\documentclass[%
    aps,
    prd,
    reprint,
    superscriptaddress,
    floatfix,
    nofootinbib,
    noamsmath,
    noamssymb
]{revtex4-1}

\usepackage[T1]{fontenc}
\usepackage[utf8]{inputenc}
\usepackage{graphicx}
\usepackage{xcolor}
\usepackage{mathtools}
\usepackage{amssymb}
\usepackage{lmodern}
\usepackage{bm}
\usepackage{siunitx}
\usepackage[final]{microtype}
\usepackage{hyperref}
\usepackage{multirow}

\newcommand{\mytitle}{QCD phase transitions in the light quark chiral limit}

\newcommand{\JLU}{%
    Institut f\"{u}r Theoretische Physik,
    Justus-Liebig-Universit\"{a}t Gie\ss{}en,
    35392 Gie\ss{}en,
    Germany
}

\newcommand{\HFHF}{%
    Helmholtz Forschungsakademie Hessen f\"{u}r FAIR (HFHF),
    GSI Helmholtzzentrum f\"{u}r Schwerionenforschung,
    Campus Gie\ss{}en,
    35392 Gie\ss{}en,
    Germany
}

\hypersetup{%
    pdftitle = {\mytitle},
    pdfauthor = {Julian Bernhardt, Christian S. Fischer},
    bookmarksopen = true,
    colorlinks = true,
    linkcolor = green!40!black,
    citecolor = blue!50!black,
    urlcolor = blue!50!black
}

\DeclareSIUnit{\MeV}{\mega\electronvolt}
\DeclareSIUnit{\GeV}{\giga\electronvolt}
\DeclareSIUnit{\fm}{\femto\meter}

\DeclareMathOperator{\Tr}{Tr}

\DeclarePairedDelimiter{\expval}{\langle}{\rangle}

\renewcommand*{\vec}[1]{\bm{#1}}
\newcommand*{\+}{\hspace*{.08335em}}

\newcommand*{\dd}{\mathrm{d}}
\newcommand*{\ii}{\mathrm{i}}

\newcommand*{\bbZ}{\mathbb{Z}}

\newcommand*{\Tc}{T_{\textup{c}}}
\newcommand*{\Nc}{N_{c}}
\newcommand*{\Nf}{N_{f}}

\newcommand*{\OO}{\textup{O}}
\newcommand*{\gs}{g}
\newcommand*{\upu}{\textup{u}}
\newcommand*{\upd}{\textup{d}}
\newcommand*{\ups}{\textup{s}}
\newcommand*{\massu}{m_{\textup{u}}}
\newcommand*{\massd}{m_{\textup{d}}}

\newcommand*{\massl}{m_{\ell}}
\newcommand*{\masss}{m_{\textup{s}}}
\newcommand*{\muu}{\mu_{\textup{u}}}
\newcommand*{\mud}{\mu_{\textup{d}}}

\newcommand*{\muB}{\mu_{\textup{B}}}

\newcommand{\pion}{\pi}
\newcommand{\kaon}{K}
\newcommand{\etames}{\eta_{8}}
\newcommand{\etapmes}{\eta_{0}}
\newcommand{\etalmes}{\eta_{\ell}}
\newcommand{\etasmes}{\eta_{\ups}}
\newcommand{\sigmes}{\sigma}
\newcommand{\fmes}{f_{0}}

\newcommand*{\SU}{\textup{SU}}
\newcommand*{\SUA}{\textup{SU}_{A}}

\newcommand*{\UA}{\textup{U}_{A}}
\newcommand*{\ZZ}{\textup{Z}}
\newcommand*{\mstri}{\masss^{\textup{tri}}}
\newcommand*{\Pps}{\textup{ps}}
\newcommand*{\Psc}{\textup{sc}}

\definecolor{dgreen}{rgb}{0.1,0.5,0.1}
\definecolor{lblue}{rgb}{0.2,0.35,1}
\definecolor{webred}{rgb}{0.75,0,0}

\begin{document}

\title{\mytitle}

\author{Julian Bernhardt}
\email{julian.bernhardt@physik.uni-giessen.de}
\affiliation{\JLU}
\affiliation{\HFHF}

\author{Christian S.~Fischer}
\email{christian.fischer@theo.physik.uni-giessen.de}
\affiliation{\JLU}
\affiliation{\HFHF}

\begin{abstract}
We investigate the order of the QCD chiral transition in the limit of vanishing
bare up/down quark masses and variations of the bare strange-quark mass $0 \le
\masss \le \infty$. In this limit and due to universality long range
correlations with the quantum numbers of pseudoscalar and scalar mesons may
dominate the physics. In order to study the interplay between the microscopic
quark and gluon degrees of freedom and the long range correlations we extend a
combination of lattice Yang--Mills theory and a (truncated) version of
Dyson--Schwinger equations by also taking back-reactions of mesonic degrees of
freedom into account. Both this system and the meson backcoupling approach have
been studied extensively in the past but this is the first work in a full $(2 +
1)$-flavor setup. Starting from the physical point, we determine the chiral
susceptibilities for decreasing up/down quark masses and find good agreement
with both lattice and functional renormalization group results. We then proceed
to determine the order of the chiral transition along the left-hand side of the
Columbia plot, for chemical potentials in the range $-(30 \,\mbox{MeV})^2 \le
\mu_q^2 \le (30 \,\mbox{MeV})^2$. We find a second-order phase transition
throughout and no trace of a first-order region in the $\Nf = 3$ corner of the
Columbia plot. This result remains unchanged when an additional Goldstone boson
due to a restored axial $\UA(1)$ is taken into account.
\end{abstract}

\maketitle

\section{\label{sec:introduction}%
    Introduction
}

It is one of the main goals of the Beam Energy Scan program at RHIC/BNL
\cite{Bzdak:2019pkr} and the ongoing and future HADES/CBM experiment at GSI/FAIR
\cite{Friman:2011zz, Salabura:2020tou, Almaalol:2022xwv} to unravel the possible
existence and location of a critical endpoint (CEP) in the chiral phase
transition line of the QCD phase diagram. The quest of extracting signals for
such a CEP from the experimental data is quite delicate and much work is
currently being invested to improve the rigorousness of theory-experiment
connections (see, e.g., \cite{Braun-Munzinger:2015hba,Luo:2017faz} for reviews).

On the theoretical side, there is wide-spread consensus on the crossover nature
of the chiral transition of QCD at zero chemical potential. The corresponding
pseudocritical temperature has been localized around $\Tc \approx
\SI{155}{\MeV}$ \cite{Borsanyi:2010bp,Bazavov:2011nk} with a couple of MeV
difference between different definitions of the chiral order parameter.
Furthermore, thermodynamic properties of the hot matter in a broad temperature
range around $\Tc$ have been determined with great accuracy
\cite{Borsanyi:2010cj,Borsanyi:2013bia,HotQCD:2014kol,Ding:2015ona,Bazavov:2017dus}.

Different theoretical approaches to QCD agree with each other that no CEP is
found in the region of the temperature--baryon-chemical-potential plane $(T,
\muB)$ with $\muB\+/\+T < 2.5$. This region is excluded by recent studies on the
lattice, see, e.g., Refs.~\cite{HotQCD:2018pds,Borsanyi:2020fev} and references
therein, as well as studies using functional methods
\cite{Fischer:2014ata,Isserstedt:2019pgx,Fu:2019hdw,Gao:2020qsj,Gao:2020fbl,Gunkel:2021oya}.
 Beyond this region, errors in lattice extrapolations accumulate rapidly and no
definite statements can be made. On the other hand, functional approaches,
i.e., approaches via Dyson--Schwinger equations (DSE) and/or the functional
renormalization group (FRG), do in principle allow for a mapping of the whole
QCD phase diagram but inherently depend on approximations and truncations
necessary to make the equations tractable.

\begin{figure*}
    \centering
    \includegraphics[scale=1.0]{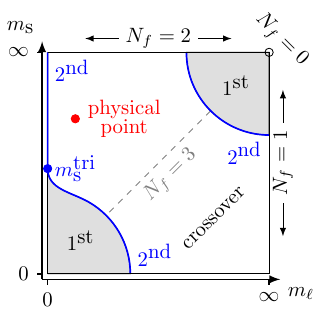}
    \hfil
    \includegraphics[scale=1.0]{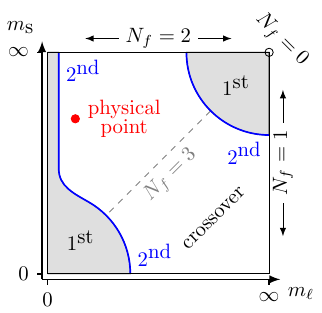}
    \hfil
    \includegraphics[scale=1.0]{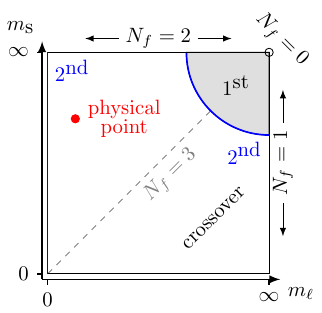}
    \caption{\label{fig:columbia-plot}%
        Three different versions of the Columbia plot \cite{Brown:1990ev} of
        phase-transition orders at nonzero temperature and vanishing chemical
        potential as functions of quark masses. The `standard plot' with
        anomalously broken $\UA(1)$-symmetry is shown on the left, in the middle
        we display a possible version with restored $\UA(1)$ and on the right we
        show an alternative version without chiral first order regions. Here,
        we assume mass-degenerate up and down quarks, $\massl = \massu =
        \massd$.%
    }
\end{figure*}

One way to learn more on the behavior of QCD for physical quark masses is to map
out the behavior at unphysical up-, down- and strange-quark masses and track
structures like critical lines and surfaces at zero, imaginary and real chemical
potential. The variation of these reveal areas of different type of transitions,
sketched in Fig.~\ref{fig:columbia-plot}, the `Columbia plot'
\cite{Brown:1990ev}. Each of these transitions is related to  an underlying
symmetry of QCD: chiral symmetry and center symmetry. Their explicit breaking
due to non-vanishing (chiral) or non-infinite (center) quark masses generates
possible patterns for the order of the transition at finite temperature and
vanishing chemical potential as a function of the quark masses displayed in
Fig.~\ref{fig:columbia-plot}. In the upper right corner of each of the three
plots, we find the first order deconfinement transition in the pure gauge limit
of infinite quark masses, separated by a second-order critical line from the
crossover region. The second-order separation line in the upper right corner of
the Columbia plot is in the $\ZZ(2)$ universality class and its location in the
$\upu$/$\upd$--$\ups$-quark-mass plane has been mapped out by lattice gauge
theory \cite{deForcrand:2010he,Saito:2011fs,Fromm:2011qi,Ejiri:2019csa,Cuteri:2020yke,Kiyohara:2021smr},
effective models \cite{Kashiwa:2012wa,Lo:2014vba}, the Dyson--Schwinger approach
\cite{Fischer:2014vxa} and background-field techniques
\cite{Reinosa:2015oua,Maelger:2017amh}. Thus, although the precise location of
the second-order critical line may differ between the approaches, the
qualitative picture is undisputed.

This is different for the chiral upper left and lower left corners of the
Columbia plot, and the left-hand side of varying strange-quark masses in the
light chiral limit. This region is governed by the chiral transition and the
corresponding axial symmetries $\UA(1) \times \SUA(\Nf)$. Whereas the latter one
is broken dynamically at low temperatures (and always explicitly by nonzero
quark masses), the former one is broken anomalously. Both the dynamical and
anomalous breaking can be restored at large temperatures, albeit the
corresponding transition temperatures may very well differ from each other.
Whether $\UA(1)$ remains broken at the chiral $\SUA(\Nf)$ transition is an open
question with conflicting indications in both directions
\cite{Brandt:2016daq,Tomiya:2016jwr,Aoki:2021qws,HotQCD:2012vvd,Buchoff:2013nra,Bhattacharya:2014ara,Dick:2015twa,Ding:2020xlj,Kaczmarek:2021ser}

The fate of the $\UA(1)$ symmetry is expected to affect the order of the chiral
$\SUA(\Nf)$ transition. With an anomalously broken $\UA(1)$ at all temperatures,
it has been conjectured that the chiral transition for the two-flavor theory
(upper left corner) is second order and in the universality class of the
$\OO(4)$ theory, whereas the chiral three-flavor theory (lower left corner) is
expected to be first order \cite{Pisarski:1983ms}, since no three-dimensional
$\SU(\Nf\le 3)$ second-order universality class is known
\cite{Butti:2003nu,deForcrand:2017cgb}. Consequently, these regions are
connected and the left-hand side of the Columbia plot features a tricritical
strange-quark mass $\mstri$ where the first-order region around the chiral
three-flavor point merges into the second-order line connected to the chiral
two-flavor point. This is the `standard' plot seen on the left of
Fig.~\ref{fig:columbia-plot}. The middle diagram of Fig.~\ref{fig:columbia-plot}
shows a possible scenario with restored $\UA(1)$. Then the upper left corner may
remain first order \cite{Pisarski:1983ms} and the two first-order corners are
expected to be connected along the left-hand side of the plot.

It is currently an open question which of these scenarios is realized in QCD.
The situation in the upper left corner and, related, in the light-quark chiral
limit of the
$\Nf=2+1$-theory with strange-quark mass fixed is not clear and indications from
lattice simulations vary between favoring either of the two left scenarios of
Fig.~\ref{fig:columbia-plot}
\cite{Iwasaki:1996ya,DElia:2004uwa,DElia:2005nmv,Kogut:2006gt,Bonati:2014kpa,Dick:2015twa,Philipsen:2016hkv,Cuteri:2017gci,Ding:2018auz}.
Both scenarios of Fig.~\ref{fig:columbia-plot} can be also realized in effective
low energy QCD models such as the PQM or PNJL model, see, e.g.,
\cite{Lenaghan:2000kr,Kovacs:2006ym,Fukushima:2008wg,Schaefer:2008hk,Mitter:2013fxa,Grahl:2013pba,Eser:2015pka,Resch:2017vjs}
 and Refs. therein and FRG approaches to QCD \cite{Braun:2009gm,Braun:2020ada}.
In Ref.~\cite{Resch:2017vjs}, it has been demonstrated that results on the
Columbia plot from mean-field approaches are substantially modified once
fluctuations have been included.

For the theory with three degenerate flavors, lattice studies support
the existence of a first-order transition for light-quark masses on coarse
lattices
\cite{Karsch:2001nf,Karsch:2003va,deForcrand:2007rq,Ding:2011du,Jin:2014hea,Takeda:2016vfj,Bazavov:2017xul}.
 However, the size of the first-order region depends strongly on the
formulation of the lattice action and the temporal extend of the lattice and
has not yet been determined unambiguously. Thus, it has been conjectured
\cite{deForcrand:2017cgb} that the third option for the Columbia plot show in
the right diagram of Fig.~\ref{fig:columbia-plot} is a realistic possibility.
Indeed, recent results on the lattice by Cuteri, Philipsen and Sciarra clearly
point in this direction \cite{Cuteri:2021ikv} and have been followed up in
\cite{Dini:2021hug} with similar results. In Ref.~\cite{Fejos:2022mso}, it has
been suggested that a second-order $\Nf=3$ transition may not be at odds with
previous FRG results, see \cite{Resch:2017vjs} and references therein.

It is the purpose of this work to re-examine the situation in a functional
continuum framework that takes both, microscopic quark and gluon degrees of
freedom and effective, long-range degrees of freedom with the quantum numbers of
pseudoscalar and scalar mesons, into account. In the framework of
Dyson--Schwinger equations (DSE) that we employ, these appear naturally as part
of fermion four-point functions in the DSE for the quark--gluon vertex
\cite{Fischer:2007ze,Fischer:2008sp}. At nonzero temperature and chemical
potential, the corresponding framework has been already explored for physical
quark masses \cite{Gunkel:2020wcl,Gunkel:2019xnh,Gunkel:2021oya} and has led to
a prediction of the location of the critical endpoint in agreement with recent
FRG studies \cite{Fu:2019hdw,Gao:2020qsj,Gao:2020fbl}. Here, quarks have been
taken into account on the $\Nf=2+1$ level but the meson sector remained $\Nf=2$
\cite{Gunkel:2020wcl,Gunkel:2019xnh,Gunkel:2021oya}. In this work, we extend the
framework to a consistent $\Nf=2+1$ level and therefore make it suitable for a
study of the left-hand side of the Columbia plot.

The paper is organized as follows. In the next section \ref{sec:framework}, we
detail the framework of Dyson--Schwinger equations including the quark and gluon
DSEs as well as the above-mentioned fluctuation effects on the quark--gluon
vertex. We discuss our treatment of the corresponding meson masses and decay
constants for varying strange-quark mass, taking particularly care of the limits
$m_s \rightarrow \infty$ and $m_s \rightarrow 0$. In Section~\ref{sec:results},
we then present our results for the order of the phase transition along the left
hand side (i.e., chiral up/down quarks but varying strange-quark mass)
of the Columbia plot for zero and small real and imaginary chemical potential.
We discuss critical temperatures, the resulting Columbia plot and the dependence
of our result on the restoration temperature of the $\UA(1)$ symmetry. We
conclude in Section~\ref{sec:summary}.

\section{\label{sec:framework}%
    Framework
}

\subsection{\label{subsec:dse}%
    Dyson--Schwinger equations
}

All necessary quantities for our investigation of the Columbia plot can be
obtained directly from dressed (i.e., full) quark propagator $S_{f}$. For a
given quark flavor $f \in \{\upu, \upd, \ups\}$, its inverse at nonzero
temperature $T$ and quark chemical potential $\mu_{f}$ is given by
\begin{equation}\label{eq:quark_propagator}
    S_{f}^{-1}(p)
    =
    \ii \gamma_{4} \tilde{\omega}_{n}^{f} C_{f}(p)
    +
    \ii \vec{\gamma} \cdot \vec{p} A_{f}(p)
    +
    B_{f}(p)
    \,.
\end{equation}
Here, $p = (\vec{p}, \tilde{\omega}_{n})$ represents the four-momentum, while
$\tilde{\omega}_{n}^{f} = \omega_{n} + \ii \mu_{f}$ denotes a combination of the
fermionic Matsubara frequencies $\omega_{n} = (2n + 1) \pi T$, $n \in \bbZ$,
with the chemical potential. All non-perturbative information such as the
non-trivial momentum dependence is carried by the quark dressing functions
$A_{f}$, $B_{f}$ and $C_{f}$.

\begin{figure}
    \centering
    \includegraphics[width=\columnwidth]{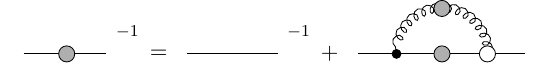}
    \\[0.25em]
    \includegraphics[width=\columnwidth]{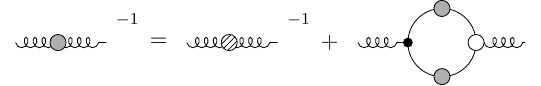}
    \caption{\label{fig:dses}%
        General form of the DSE for the quark propagator (top) and truncated
        gluon DSE (bottom). Large grey and white dots indicate dressed
        quantities; solid and curly lines represent quark and gluon propagators,
        respectively. There is a separate quark DSE for the up, down and for
        the strange quarks. The large shaded dot denotes the quenched gluon
        propagator that is taken from the lattice while the quark loop is
        evaluated explicitly. The latter contains an implicit flavor sum over
        up, down and strange.
    }%
\end{figure}

To calculate the quark propagator, we solve its associated Dyson--Schwinger
equation (DSE) which reads
\begin{equation}
    \label{eq:quark_dse}
    S_{f}^{-1}(p)
    =
    Z_{2}
    \bigl(
        \ii \gamma_{4} \tilde{\omega}_{n}^{f}
        +
        \ii \vec{\gamma} \cdot \vec{p}
        +
        Z_{m} m_{f}
    \bigr)
    -
    \Sigma_{f}(p)
    \,.
\end{equation}
Above, $m_{f}$ denotes the current-quark mass while $Z_{2}$ and $Z_{m}$ labeling
the wave function and mass renormalization constants, respectively, which are
calculated in vacuum using a momentum-subtraction scheme.

\begin{figure*}
    \centering
    \includegraphics[scale=1.0]{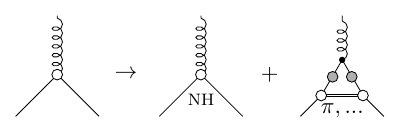}
    \\[0.25em]
    \includegraphics[scale=1.0]{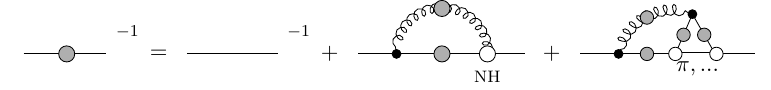}
    \caption{\label{fig:vertex-ansatz}%
        \emph{Top}: Vertex ansatz for including long-range correlations
        originating the in the skeleton expansion of the quark--gluon-vertex
        DSE. \emph{Bottom}: Resulting DSE for the quark propagator.%
    }
\end{figure*}

The quark DSE is displayed pictorially in the top row of Fig.~\ref{fig:dses}.
The quark self-energy $\Sigma_{f}$ comprises both the gluon propagator and the
quark--gluon vertex. In our framework, we calculate the gluon propagator
explicitly by solving its DSE albeit in a truncated form (for details see
below). For the quark--gluon vertex we rely on a well-tested vertex model
constructed to (approximately) satisfy Slavnov--Taylor identities and preserving
all perturbative and renormalization constraints (see below)
\cite{Fischer:2012vc,Fischer:2014ata,Eichmann:2015kfa,Isserstedt:2019pgx,Isserstedt:2020qll}.

As the correlation length diverges in the vicinity of a second-order phase
transition, long-range correlations in the quark--gluon vertex become important.
These arise from a specific diagram in the DSE for the quark--gluon vertex that
involves a four-quark kernel. In pole approximation, this diagram is shown in
the upper equation of Fig.~\ref{fig:vertex-ansatz}. This diagram provides
contributions to all tensor components of the quark--gluon vertex
\cite{Fischer:2007ze}. In the quark DSE, the resulting two-loop diagram can be
simplified to a one-loop diagram using a homogenous BSE, see
Refs.~\cite{Fischer:2007ze,Gunkel:2021oya} for details. The effect of this
specific contribution to the quark--gluon interaction has been studied in a
number of works at zero temperature/chemical potential including a discussion of
the analytic structure of the quark propagator \cite{Fischer:2008sp}, a
discussion of its effect onto the meson spectrum \cite{Fischer:2008wy}, and an
exploratory study of meson-cloud effects in baryons
\cite{Sanchis-Alepuz:2014wea}. In all these studies, it has been noted that
meson-backcoupling effects typically provide contributions of the order of
10--20\,\% as compared with other components of the quark--gluon interaction.
The effect of this contribution on the location of the CEP has been studied in
Ref.~\cite{Gunkel:2021oya}.

The quark--gluon vertex, split into a non-hadronic (NH) and a mesonic (M) part
and inserted into the quark-DSE leads to the following expression for the
analogous splitting of the quark self-energy:
\begin{equation}
    \Sigma_{f}(p)
    =
    \Sigma^{\text{NH}}_{f}(p)
    +
    \Sigma^{\text{M}}_{f}(p)
    \,,
\end{equation}
The non-hadronic part of the quark self-energy corresponds to the usual quark
self-energy from previous works:
\begin{multline}\label{eq:quark_self-energy}
    \Sigma^{\text{NH}}_{f}(p)
    =
    (\ii \gs)^{2}
    \frac{4}{3}
    \frac{Z_{2}}{\tilde{Z}_{3}}
    T
    \sum_{\omega_{n}}
    \int \frac{\dd^{3} q}{(2 \pi)^{3}}
    D_{\nu\rho}(k)
    \gamma_{\nu}
    \times
    \\
    \times
    S_{f}(q)
    \Gamma_{\rho}^{f}(p, q; k)
    \,.
\end{multline}
Here, $k = p - q$ indicates the gluon momentum, $g$ labels the strong coupling
constant, $\tilde{Z}_{3}$ represents the ghost renormalization constant and
$D_{\nu\rho}$ is the dressed gluon propagator. The prefactor of $4/3$ originates
in the trace over color space which has already been carried out for $\Nc = 3$
colors.

The new element of the truncation used in this work as compared to
Ref.~\cite{Gunkel:2021oya} is the inclusion of the strange-quark contributions
to the mesonic part of the vertex. The necessary technical steps are discussed
in detail in Subsection~\ref{subsec:backcoupling} below.

In Eq.~\eqref{eq:quark_self-energy}, $\Gamma_{\rho}^{f}$ labels the non-hadronic
quark--gluon vertex for which we employ the ansatz:
\begin{multline}
    \Gamma^{f}_{\rho}(p, q; k)
    =
    \Gamma(x)
    \biggl[
        \delta_{\rho i}
        \gamma^{i}
        \frac{A_{f}(p) + A_{f}(q)}{2}
        +
        \\
        +
        \delta_{\rho 4}
        \gamma^{4}
        \frac{C_{f}(p) + C_{f}(q)}{2}
    \biggr]
    \,.
\end{multline}
This is a combination of the leading Dirac tensor structure of the Ball--Chiu
vertex \cite{Ball:1980ay} -- which leads to backcoupling effects of the quarks
onto the vertex -- with a phenomenological vertex dressing function:
\begin{equation}\label{d1}
    \Gamma(x)
    =
    \frac{d_{1}}{d_{2} + x}
    +
    \frac{x}{\Lambda^{2} + x}
    \biggl(
        \frac{\alpha \beta_{0}}{4 \pi}
        \ln\bigl(
            x / \Lambda^{2} + 1
        \bigr)
    \biggr)^{2 \delta}
    \,.
\end{equation}
The first term is an IR enhancement inspired by Slavnov--Taylor identities while
the second term ensures the correct perturbative behavior of the propagators in
the UV. The running coupling is given by $\alpha = 0.3$, the scales $d_{2} =
\SI{0.5}{\GeV^{2}}$ and $\Lambda = \SI{1.4}{\GeV}$ are fixed to match the ones
in the gluon lattice data. The anomalous dimension reads $\delta = -9 \Nc /
(44\Nc - 8 \Nf)$ and $\beta_{0} = (11 \Nc - 2 \Nf) / 3$. For the argument, we
have $x = k^{2}$ in the quark DSE while $x = p^{2} + q^{2}$ in the gluon DSE in
Eq.~\eqref{eq:quark_loop} to ensure multiplicative renormalizability. More
details on the vertex and the choice of the parameters can be found in
Refs.~\cite{Fischer:2014ata,Isserstedt:2019pgx} and references therein.

Since the meson-backcoupling diagrams originate in a modification of the vertex,
we need to adjust the vertex-strength parameter $d_{1}$. We tune $d_{1}$ such
that the pseudocritical temperature obtained from the chiral susceptibility at
the physical point, i.e., the maximum of $m_{\pi} = \SI{139}{\MeV}$ in
Fig.~\ref{fig:msp-susc}, corresponds to the one from the lattice $\Tc^{p} =
\SI{156.5}{\MeV}$. This yields $d_{1} = \SI{8.98}{\GeV^{2}}$ as opposed to
$d_{1} = \SI{8.49}{\GeV^{2}}$ without meson contributions
\cite{Isserstedt:2019pgx}.

The gluonic part of our truncation is unchanged as compared to previous works,
i.e., in the Yang--Mills sector we do not take the meson effects explicitly into
account\footnote{%
    The main reason is feasibility: in the quark-loop diagram of the gluon-DSE
    the meson-exchange diagram remains two-loop and is therefore too expensive
    in terms of CPU time. However, these contributions are also irrelevant when
    it comes to critical exponents \cite{Fischer:2011pk}.%
}. As stated earlier, the gluon propagator is calculated using a simplified
version of the full gluon DSE (illustrated in the bottom row of
Fig.~\ref{fig:dses}):
\begin{equation}\label{eq:gluon_dse}
    D_{\nu\rho}^{-1}(k)
    =
    \bigl[
        D_{\nu\rho}^{\textrm{YM}}(k)
    \bigr]^{-1}
    +
    \Pi_{\nu\rho}(k)
    \,.
\end{equation}
Here, $D_{\nu\rho}^{\textrm{YM}}$ denotes the quenched gluon propagator given by
a combination of all pure Yang--Mills for which we use temperature-dependent
fits to results of quenched lattice calculations
\cite{Fischer:2010fx,Maas:2011ez,Eichmann:2015kfa} as input. The quark loop
$\Pi_{\nu\rho}$, however, is calculated explicitly within our framework. This
leads to an unquenching of the gluon and consequently a non-trivial backcoupling
of the chiral dynamics of the quarks onto the gluon. It is given by
\begin{multline}\label{eq:quark_loop}
    \Pi_{\nu\rho}(k)
    =
    \frac{(\ii \gs)^{2}}{2}
    \sum_{f}
    \frac{Z_{2}}{\tilde{Z}_{3}}
    T
    \sum_{\omega_{n}}
    \int \frac{\dd^{3} q}{(2 \pi)^{3}}
    \Tr\bigl[
        \gamma_{\nu}
        S_{f}(q)
        \times
        \\
        \times
        \Gamma_{\rho}^{f}(p, q; k)
        S_{f}(p)
    \bigr]
    \,.
\end{multline}
The prefactor of $1/2$ stems from the color trace, the flavor sum $f \in
\{\upu/\upd, \ups\}$ runs over the investigated $2+1$ quark flavors. We work in
the isospin-symmetric limit of degenerate up and down quarks ($\massu = \massd$,
$\muu = \mud$). Therefore, we will from now on only refer to a `light' quark
$\ell = \upu = \upd$ for the sake of simplicity. At the physical point, the
quark masses are fixed using results for the pion and kaon masses in vacuum
obtained from the Bethe-Salpeter formalism developed in
Ref.~\cite{Heupel:2014ina}. This leads to values of $\massl = \SI{0.8}{\MeV}$
and $\masss = \SI{20.56}{\MeV}$ at a renormalization point of $\SI{80}{\GeV}$.

\subsection{\label{subsec:backcoupling}%
    Meson-backcoupling self-energy
}

\begin{figure}
    \centering
    \includegraphics[scale=1.0]{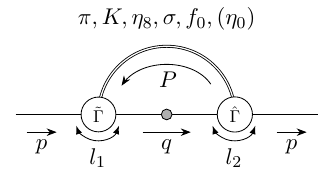}
    \caption{\label{fig:mesonbc-selfenergy}%
        One-loop meson-backcoupling diagram in the quark self-energy as an
        approximation of the two-loop diagram of Fig.~\ref{fig:vertex-ansatz}.%
    }
\end{figure}

In this subsection, we discuss the mesonic part of the quark self-energy in some
detail. To this end, we begin again with the two-loop expression in the quark
DSE in the lower equation of Fig.~\ref{fig:vertex-ansatz}. This diagram
originates from the meson pole approximation of a fermion four-point function in
the DSE for the quark--gluon vertex \cite{Fischer:2007ze}. The corresponding
meson propagator in the diagram is therefore bare and accompanied by two
Bethe--Salpeter amplitudes that connect the quark lines with the exchanged meson
in question. In Ref.~\cite{Fischer:2007ze}, it has been realized that the left
half of the two-loop diagram displayed in Fig.~\ref{fig:vertex-ansatz} can be
interpreted as the interaction diagram in a homogeneous Bethe--Salpeter equation
(BSE) and therefore can be replaced with a Bethe--Salpeter amplitude (BSA). This
way, the mesonic part of the quark self-energy reduces to a one-loop diagram
illustrated in Fig.~\ref{fig:mesonbc-selfenergy}. Of course, this simplifies
calculations tremendously.

In principle, this diagram contains mesons with all quantum numbers that can be
build from a quark--antiquark pair. In practice, we are only interested in those
mesons that have the potential to become massless at phase transitions, i.e.,
the pseudoscalar meson octet, its critical chiral partner modes and the
pseudoscalar singlet in case the axial $\UA(1)$ is restored at the transition
temperature. All these are potentially long ranged and are expected to become
the dominant degrees of freedom at second-order phase transitions. All other
meson contributions are subleading due to their large masses in the meson
propagator and are therefore omitted in our approach.

We thus end up with the lightest pseudoscalar octet, i.e., pions, kaons and the
$\eta_{8}$, as well as the $\eta_{0}$ in a crosscheck calculation (see
Section~\ref{sec:results}). Additionally, we consider the scalar $\sigma$ meson
(i.e., the $f_{0}(500)$) as it is vital for the correct $\OO(4)$-scaling
behavior in the upper left corner of the Columbia plot and the $\ups
\bar{\ups}$-partner of the $\sigma$ (which we identify with the $f_{0}(980)$)
that may be important in the $\Nf = 3$ chiral limit. As detailed below,
Eq.~(\ref{masses}), we assume the $f_{0}$ to be massless in the chiral $\Nf=3$
limit, since it has the quantum numbers of the strange-quark condensate. In
order to obtain a consistent $\Nf = 3$ limit, we alter its flavor factor by hand
to match the one of the $\sigma$ meson thus obtaining three identical DSEs for
the up, the down and the strange quark in this limit (cf.
Tab.~\ref{tab:meson-info}).\footnote{%
    In a more complete framework, we would additionally include the $a_{0}$ and
    all other members of the scalar meson multiplets and determine their masses,
    wave functions and decay constants dynamically. In this case, adjusting
    flavor factors by hand would not be necessary.%
}

Restricting to the pions and $\sigma$ meson, this type of meson backcoupling was
discussed in detail, e.g., in Refs.~\cite{Fischer:2011pk, Gunkel:2021oya}.
Building on the explanations therein, we generalize this to the $\Nf=2+1$ case
to arrive at the following mesonic contribution to the quark self-energy:
\begin{multline}\label{eq:mesonbc}
    \Sigma^{\mathrm{M}}_{f}(p)
    =
    \sum_{X} F^{f}_{X}
    \sum_{\omega_{q}} \int \frac{\dd^{3} q}{(2 \pi)^{3}}
    D_{X}(P) \times
    \\
    \times
    \tilde{\Gamma}^{f}_{X}(l_{1}, -P) \,
    S^{f}_{X}(q) \,
    \hat{\Gamma}^{f}_{X}(l_{2}, P)
    \,.
\end{multline}
Here, $P = p - q$ denotes the meson's total momentum while $l_{i}$ represent the
relative momenta of the Bethe--Salpeter amplitudes. Using an appropriate
momentum routing, we can identify $l_{1} = p$ and $l_{2} = q$.

Most quantities in Eq.~\eqref{eq:mesonbc} are specific to the flavor of the
external quark $f \in \{\ell, \ups\}$ and the exchanged meson $X \in
\{\pion,\kaon, \etames, \sigmes, \fmes, (\eta_0)\}$. First, there are the
multiplicities $F^{f}_{X}$ of the respective meson-backcoupling diagram obtained
by performing the trace over flavor space. Second, we have the quark propagator
of the internal quark $S^{f}_{X}$ which differs from the external one for the
kaon, i.e., $S^{\ell}_{\kaon} = S_{\ups}$ and $S^{\ups}_{\kaon} = S_{\ell}$.

The (inverse) meson propagator at nonzero temperature is given by
\cite{Son:2001ff}:
\begin{equation}
    D^{-1}_{X}(P)
    =
    P_{4}^{2} + u_{X}^{2} \big(\vec{P}^{2} + m_{X}^{2}\big)
    \,,
\end{equation}
with $u_{X} = f^{s}_{X} / f^{t}_{X}$ being the meson velocity which is given by
the ratio of the spatial and temporal meson decay constants, $f^{s}_{X}$ and
$f^{t}_{X}$, respectively. Again, we restrict ourselves to potentially critical
modes and consider only the zeroth Matsubara frequency of the meson propagator,
i.e., $P_{4} = 0$. All other Matsubara frequencies act as an effective meson
mass that leads to suppression of the respective contribution. This restriction
implies $\omega_{q} = \omega_{p}$ in Eq.~\eqref{eq:mesonbc} and consequently
cancels the Matsubara sum \cite{Maris:2000ig}.

For the meson masses $m_{X}$, we choose the vacuum values
\begin{align}\label{masses}
\begin{split}
    m_{\pion}
    &=
    \SI{156.525}{\MeV^{1/2}}
    \cdot
    \sqrt{\massu}
    \,,
    \quad
    m_{\sigma}
    =
    2 m_{\pion}
    \,,
    \\
    m_{\kaon}
    &=
    \SI{74.2}{\MeV^{1/2}}
    \cdot
    \sqrt{\masss}
    +
    1.54
    \cdot
    \masss
    \,,
    \\
    m_{\etames}
    &=
    m_{\fmes}
    =
    2 m_{\kaon}
    \,.
\end{split}
\end{align}
that have been obtained in the following way: We have solved the coupled system
of $\Nf=2+1$ DSEs without meson-backreaction effects in the vacuum and for
complex quark momenta and then extracted the corresponding meson masses from
solving their corresponding Bethe--Salpeter equations for different up/down- and
strange-quark masses. The resulting mass curves for the pion and the kaon have
been fitted with the expressions above, which correspond to the expected
behavior from Gell-Mann--Oakes--Renner relations. The remaining masses are
expressed in terms of these for the sake of simplicity in such a fashion that
the correct massless modes appear in the $\Nf=2$ and $\Nf=3$ chiral flavor
limits.

Note that this treatment overestimates the effects of the critical modes in the
low-temperature, chirally broken phase since the critical modes are already
massless by construction instead of becoming massless at the critical
temperature. We have checked that this simplification does not affect the order
of the transition, but it may affect the location of the transition, i.e., the
critical temperature. We discuss this further in Section~\ref{sec:results}, when
we present our results. In principle, one could solve the
temperature-dependent Bethe--Salpeter equations also at nonzero temperature
along the lines of Ref.~\cite{Maris:2000ig}. There, it has been shown explicitly
that the pion and sigma modes follow the correct pattern of symmetry breaking
and restoration in the $\Nf=2$ chiral limit. In practice, this would add an
extra layer of complications and an order of magnitude more in computing time to
an already demanding endeavor and we there resort to the simplifications
expressed in Eq.~(\ref{masses}).

The central unknown quantities are the meson Bethe--Salpeter amplitudes
$\hat{\Gamma}_{X}$. In the chiral limit, it is an exact property of QCD
\cite{Maris:1997tm,Maris:1997hd} that the leading BSA of the Goldstone boson can
be expressed through the scalar dressing function $B$ of the quark propagator
and the Goldstone-boson decay constant via $\Gamma_{X}(l) = \gamma_5
B(l)/f_{X}$, with relative momentum $l$ between the quark and the antiquark, see
\cite{Eichmann:2016yit} for a review and a detailed explanation of this
property. This behavior persists approximately also away from the chiral limit
with the caveat that the quark dressing function then develops a logarithmic
tail at large momenta, whereas the Bethe--Salpeter amplitude always falls like a
power in momentum. We therefore adopt the following prescription for our meson
amplitudes:
\begin{equation}\label{eq:goldberger-treiman}
    \hat{\Gamma}_{X}^{f}(l, P)
    =
    \gamma_{X} \frac{B_{f}(l)}{f^{f, t}_{X}}
    \cdot
    \frac{a}{a + l^{2}}
    \,,
    \quad
    a = \SI{80}{\GeV^{2}}
    \,,
\end{equation}
where $f^{f, t}_{X}$ labels the respective temporal meson decay constant and
$\gamma_{X} = \gamma_{5}$ for the pseudoscalar mesons and $\gamma_{X} = I$ for
the scalar mesons. Additionally, we also apply these relations to mesons
comprising non-chiral strange quarks. To account for the correct power-law
behavior in the large momentum limit, we supplement the quark dressing function
with a Pauli--Villars-like term with a scale that matches our renormalization
point \cite{Lucker:2013uya}. As a consequence, this also renders the
meson-backcoupling diagrams ultraviolet finite so that no further regularization
is necessary. Note that for mesons with mixed flavor content we always use the
$B$ function of the quark external to the loop in which the BSA appears. This
turned out to be numerically advantageous for a consistent $\Nf = 3$ limit.

Additionally, we introduce the closely related quantity $\tilde{\Gamma}_{X}$
which originates in the two-loop diagram of the vertex expansion. Apart from a
non-trivial sign which is a result of its two-loop origin, we identify it with
the BSA $\hat{\Gamma}_{X}$:
\begin{equation}
    \tilde{\Gamma}_{X}^{f}(l, P)
    =
    (-1)^{X} \, \hat{\Gamma}_{X}^{f}(l, P)
    \,,
\end{equation}
where $(-1)^{X} = -1$ for the pseudoscalar mesons and $(-1)^{X} = +1$ for the
scalar mesons.

\begin{figure}
    \centering
    \includegraphics[scale=1.0]{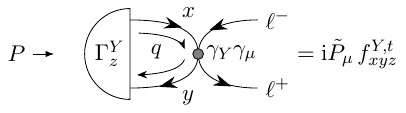}
    \caption{\label{fig:pagels}%
        Pictorial representation of the generalized Pagels--Stokar formula in
        Eq.~\eqref{eq:pagels-stokar-actual} we use to calculate the
        pseudoscalar and scalar meson decay constants.%
    }
\end{figure}

Last, we also require the meson decay constants. These are calculated using a generalized Pagels--Stokar formula \cite{Pagels:1979hd},
\begin{multline}\label{eq:pagels-stokar-actual}
    \ii \tilde{P}_{\mu} \bigl( f_{xyz}^{Y, t} \bigr)^{2}
    =
    3 T
    \sum_{\omega_{q}} \int \frac{\dd^{3} q}{(2 \pi)^{3}}
    \Tr\biggl[
    S_{x}(q + P)
    \gamma_{Y} \gamma_{\mu}
    \times
    \\
    \times
    S_{y}(q)
    \gamma_{Y}
    B_{z}(q)
    \cdot
    \frac{a}{a + q^{2}}
    \biggr]
    \,,
\end{multline}
which merits some explanations. First, we use the abbreviation $\tilde{P}_{\mu}
= (\omega_{P}, u_{X} \vec{P})$, while $x$ and $y$ label the flavor indices of
the quarks contributing to the meson in question. Consequently, the temporal and
spatial decay constants are obtained in the limit $P_{\mu} \to 0$ from the
temporal and spatial momentum component, respectively. The index $z$ labels the
quark flavor of the external quark of the backcoupling diagram where the meson
appears and $Y \in \{\text{ps}, \text{sc}\}$ represents the parity of the meson.
This relation is pictorially displayed in Fig.~\ref{fig:pagels}. This notation
is necessary for the following reason: the decay constants in our diagrams do
not depend on the mesons directly but rather on the contributing quark
propagators of the backcoupling diagram. Furthermore, they have to match the
type of scalar dressing function used in the BSA. For hidden-flavor mesons this
is unproblematic and Eq.~(\ref{eq:pagels-stokar-actual}) matches the usual
Pagels--Stokar approximations. For open-flavor mesons, however, the four kaons
in our approach, this has the consequence that the decay constant appearing in
the light-quark DSE is different from the one in the strange-quark DSE since the
external quark is different. Eq.~(\ref{eq:pagels-stokar-actual}) accounts for
this. Furthermore, we symmetrize Eq.~\eqref{eq:pagels-stokar-actual} in a
mathematically well-defined manner\footnote{%
    This is done by a shift of the integration momentum which can be compensated
    in the BSA with a different momentum routing.%
} with the arithmetic mean of the exchanged quark flavors. This way, we arrive
at the following definitions:
\begin{align}
    f_{xy}^{Y, s} f_{xy}^{X, t}
    &=
    \bigl(
        f_{xyx}^{Y, s} f_{xyx}^{Y, t}
        +
        f_{yxx}^{Y, s} f_{yxx}^{Y, t}
    \bigr) / 2
    \,,
    \\[0.25em]
    \bigl( f_{xy}^{Y, t} \bigr)^{2}
    &=
    \bigl(
        \bigl( f_{xyx}^{Y, t} \bigr)^{2}
        +
        \bigl( f_{yxx}^{Y, t} \bigr)^{2}
    \bigr) / 2
    \,,
\end{align}
from which we can obtain the required (temporal) meson decay constants used in
our calculations:
\begin{equation}
    f^{Y}_{xy}
    :=
    f^{Y, t}_{xy}
    =
    f_{xy}^{Y, s} f_{xy}^{Y, t}
    /
    \sqrt{\bigl( f_{xy}^{Y, t} \bigr)^{2}}
    \,.
\end{equation}

\begin{table}
    \centering
    \setlength{\tabcolsep}{4pt}
    \def\arraystretch{1.25}
    \begin{minipage}[t]{0.225\textwidth}
        $\Nf = 2 + 1$:\\[0.25em]
        \begin{tabular}{|c|c||c|c|c|}
            \hline
            $f$ & $X$ & $F_{X}^{f}$ & $S_{X}^{f}$ & $f_{X}^{f}$ \\
            \hline
            $\ell$ & $\pion$   & $3 / 2$ & $S_{\ell}$ & $f_{\ell\ell}^{\Pps}$ \\
            $\ell$ & $\sigma$  & $1 / 2$ & $S_{\ell}$ & $f_{\ell\ell}^{\Psc}$ \\
            $\ell$ & $\kaon$   & $1$     & $S_{\ups}$ & $f_{\ell\ups}^{\Pps}$ \\
            $\ell$ & $\etames$ & $1 / 6$ & $S_{\ell}$ & $f_{\ell\ell}^{\Pps}$ \\
            $\ell$ & $(\etapmes)$& $1/3$ & $S_{\ell}$ & $f_{\ell\ell}^{\Pps}$ \\
            $\ups$ & $\kaon$   & $2$     & $S_{\ell}$ & $f_{\ups\ell}^{\Pps}$ \\
            $\ups$ & $\etames$ & $2 / 3$ & $S_{\ups}$ & $f_{\ups\ups}^{\Pps}$ \\
            $\ups$ & $(\etapmes)$& $1/3$ & $S_{\ups}$ & $f_{\ups\ups}^{\Pps}$ \\
            $\ups$ & $\fmes$   & $1 / 2$ & $S_{\ups}$ & $f_{\ups\ups}^{\Psc}$ \\
            \hline
        \end{tabular}
    \end{minipage}
    \begin{minipage}[t]{0.225\textwidth}
        $\Nf = 2$:\\[0.25em]
        \begin{tabular}{|c|c||c|c|c|}
            \hline
            $f$ & $X$ & $F_{X}^{f}$ & $S_{X}^{f}$ & $f_{X}^{f}$ \\
            \hline
            $\ell$ & $\pion$   & $3 / 2$ & $S_{\ell}$ & $f_{\ell\ell}^{\Pps}$ \\
            $\ell$ & $\sigma$  & $1 / 2$ & $S_{\ell}$ & $f_{\ell\ell}^{\Psc}$ \\
            $\ell$ & $(\etapmes)$& $1/3$ & $S_{\ell}$ & $f_{\ell\ell}^{\Pps}$ \\
            \hline
        \end{tabular}
        \\[1em]
        $\Nf = 3$:\\[0.25em]
        \begin{tabular}{|c|c||c|c|c|}
            \hline
            $f$ & $X$ & $F_{X}^{f}$ & $S_{X}^{f}$ & $f_{X}^{f}$ \\
            \hline
            $\ell$ & $\pion,\kaon,\eta_8$   & $8 / 3$ & $S_{\ell}$ & $f_{\ell\ell}^{\Pps}$ \\
            $\ell$ & $\sigma$  & $1 / 2$ & $S_{\ell}$ & $f_{\ell\ell}^{\Psc}$ \\
            $\ell$ & $(\etapmes)$& $1/3$ & $S_{\ell}$ & $f_{\ell\ell}^{\Pps}$ \\
            \hline
        \end{tabular}
    \end{minipage}
    \caption{\label{tab:meson-info}%
        Information of multiplicities, internal quark propagators and decay
        constants for all considered meson backcoupling diagrams.%
    }
\end{table}

Of course, this procedure is only relevant for finite, nonzero strange-quark
masses along the left-hand side of the Columbia plot. In the chiral $\Nf=2$ and
$\Nf=3$ limits (lower and upper left corners), it becomes immaterial. In total,
we summarize all necessary information for our meson-backcoupling procedure
compactly in Tab.~\ref{tab:meson-info}.

\section{\label{sec:results}%
    Results and discussion
}

\begin{figure*}
    \centering
    \includegraphics[scale=1.0]{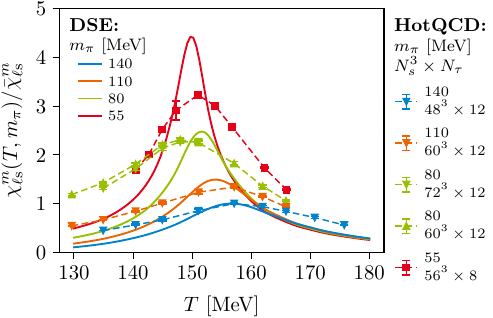}
    \hfil
    \includegraphics[scale=1.0]{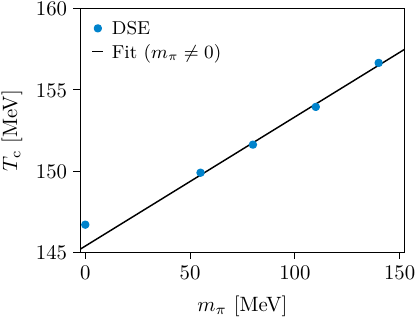}
    \caption{\label{fig:msp-susc}%
        Left: Chiral susceptibility as a function of temperature for fixed,
        physical strange-quark
        mass but different up/down-quark masses corresponding to different pion
        masses. The susceptibilities are normalized to the maximal value of
        $m_{\pi} = \SI{140}{\MeV}$.
        The lattice data have been taken from Ref.~\cite{HotQCD:2019xnw}.
        Right: Critical temperatures for the same pion masses. Shown in a
        linear extrapolation of our
        data at finite pion masses to zero mass compared with the result of an
        explicit calculation at
        zero pion mass.
    }
\end{figure*}

In this section, we present the numerical results obtained in our framework. Our
investigation of the Columbia plot is based on monitoring the behavior of the
light-quark condensate as the order parameter for chiral symmetry breaking. It
can be obtained from the quark propagator via the relation
\begin{equation}\label{eq:condensate}
    \expval{\bar{\psi} \psi}_{f}
    =
    -3 Z_{2}^{f} Z_{m_{f}} T
    \sum_{\omega_{q}} \int \frac{\dd^{3} q}{(2 \pi)^{3}}
    \Tr\bigl[ S_f(q) \bigr]
    \,,
\end{equation}
where the factor three stems from the color trace and, as always, $f \in
\{\ell, \ups\}$. The quark condensate is quadratically divergent for
all flavors with a nonzero bare-quark mass due to a contribution proportional
to $m_{f} \Lambda^{2}$ where $\Lambda$ denotes the ultraviolet (UV) momentum
cutoff. As a consequence, the divergent part of the light-quark condensate can
be cancelled with the corresponding one of the strange-quark condensate when the
latter is multiplied with the light-to-strange mass ratio. In our $(2 +
1)$-flavor setup, a regularized expression for the light-quark condensate can
therefore be obtained by considering the difference
\begin{equation}\label{eq:condensate_subtracted}
    \Delta_{\ell\ups}
    =
    \expval{\bar{\psi} \psi}_{\ell}
    -
    \frac{Z_{m}^{\ell}}{Z_{m}^{\ups}}
    \frac{m_{\ell}}{m_{\ups}} \+
    \expval{\bar{\psi} \psi}_{\ups}
    \,,
\end{equation}
Note that we are working with renormalized quantities, hence the appearance of
the renormalization constants $Z_{m}^{f}$ in order to preserve multiplicative
renormalizability. In the case of massless light quarks, the subtracted
condensate reduces to the unsubtracted one which is then well-defined, i.e.,
UV finite.

The chiral susceptibility is then defined as the derivative of the regularized
condensate with respect to the light-quark mass:
\begin{equation}
    \chi_{\ell\ups}^{m}
    =
    \frac{\partial}{\partial \massl}
    \Delta_{\ell\ups}
    \,.
\end{equation}
Up to normalization factors and for $\massl = \massu = \massd$, this definition
is equivalent to the ones used in Refs.~\cite{HotQCD:2019xnw,Braun:2020ada}.

The remainder of this section is structured as follows. First, we study the line
between the physical point and the left-hand side of the Columbia plot by
analyzing the dependence of the chiral susceptibility on the pion (and thus the
light-quark) mass. Second, we investigate the type of the chiral phase
transition across the whole left-hand side of the Columbia plot, i.e., for
chiral light quarks and strange-quark masses between $\masss \in [0, \infty)$.
We also quantify the dependence of the critical temperature on the strange-quark
mass. Third, we analyze the scaling behavior of the light-quark condensate.
Last, we extend our analysis to small but nonzero chemical potential, both real
and imaginary.

\subsection{\label{subsec:to-cl}%
    Towards the chiral limit
}
\begin{table*}
    \centering
    \def\arraystretch{1.5}
    \begin{tabular}{|c|c|c|c|c|c|c|}
        \hline
         $m_\pi$ [MeV]         &                      & $0$   &
         $55$                  & $80$                  & $110$
         &
         $140$\\
        \hline
        \multirow{5}{*}{$\Tc \ [\SI{}{MeV}]$} & DSE   &$146.7$&
        $149.9$               & $151.6$               & $154.0$               &
        $156.7$ \\
        \cline{2-7}
        & FRG~\cite{Braun:2020ada}                & $142$ &
        $148.0$               & $150.5$               & $153.6$               &
        $156.3$                \\
        \cline{2-7}
        & FRG--DSE~\cite{Gao:2021vsf}                & $141.3$ &
        $146.5$               & $149.1$               & $152.1$               &
        $155.4$                \\
        \cline{2-7}
        & HotQCD ($N_\tau = 12$)~\cite{HotQCD:2019xnw}& -     &
        -                    & $149.7^{+0.3}_{-0.3}$ & $155.6^{+0.6}_{-0.6}$ &
        $158.2^{+0.5}_{-0.5}$  \\
        \cline{2-7}
        & HotQCD ($N_\tau = 8$)~\cite{HotQCD:2019xnw} & -     &
        $150.9^{+0.4}_{-0.4}$ & $153.9^{+0.3}_{-0.3}$ & $157.9^{+0.3}_{-0.3}$ &
        $161.0^{+0.1}_{-0.1}$  \\
        \hline
    \end{tabular}
    \caption{\label{tab:tc-ontheway}%
        Comparison of critical temperatures for different up/down-quark masses
        corresponding to different pion masses and fixed physical strange-quark
        masses between our DSE findings, the FRG, FRG--DSE and the lattice
        results, respectively.%
    }
\end{table*}

We start our investigation of the Columbia plot with the line between the
physical point and the left-hand side. That is, we keep the strange-quark mass
physical $\masss = \masss^{p}$ and decrease the light-quark mass from its
physical value down to zero. This path has been explored already by the HotQCD
collaboration \cite{HotQCD:2019xnw} with lattice QCD methods, the fQCD
collaboration using the FRG~\cite{Braun:2020ada} as well as with a combined
FRG--DSE approach in Ref.~\cite{Gao:2021vsf}.

\begin{figure*}
    \centering
    \includegraphics[scale=1.0]{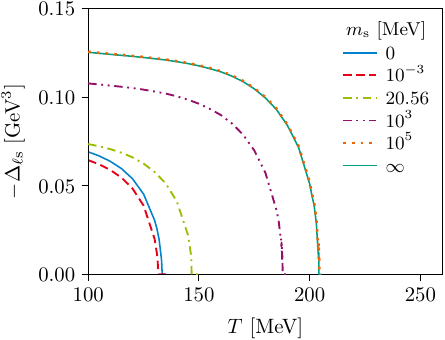}
    \hfil
    \includegraphics[scale=1.0]{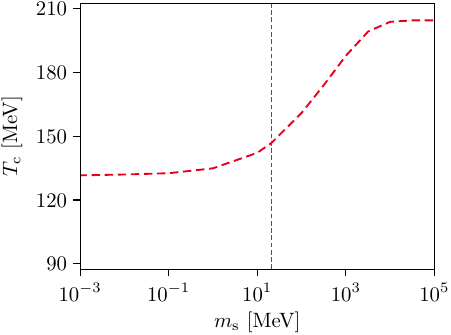}
    \caption{\label{fig:mu0-Tc-vs-ms}%
        Left: Regularized quark condensate as a function of temperature for
        different
        strange-quark masses in the up/down quark chiral limit. %
        Right: Dependence of the critical temperature on the strange-quark mass
        in the
        same limit. The dashed vertical line indicates the physical
        strange-quark mass.%
    }
\end{figure*}

In the left diagram of Fig.~\ref{fig:msp-susc}, we show the chiral
susceptibility as a function of temperature for four different pion masses
compared to the lattice results of Ref.~\cite{HotQCD:2019xnw}. Analogously to
Ref.~\cite{Braun:2020ada}, we normalize the susceptibilities to the maximal
value at a physical pion mass:
\begin{equation}
    \bar{\chi}_{\ell\ups}^{m}
    =
    -
    \max_{T} \bigl| \chi_{\ell\ups}^{m}(T, m_{\pi} = \SI{140}{\MeV}) \bigr|
    \,.
\end{equation}

Qualitatively, we find similar results as both the lattice, the FRG and the
FRG--DSE approach in Refs.~\cite{HotQCD:2019xnw,Braun:2020ada,Gao:2021vsf}. That
is, for decreasing pion masses, the peak of the susceptibilities increases in
height and moves towards lower temperatures monotonically. Quantitatively, the
results are also very similar. Namely, the peak heights are comparable with the
FRG and FRG--DSE findings for all investigated pion masses and the
(pseudo)critical temperatures away from the chiral limit do not deviate more
than $\SI{2}{\MeV}$ and $\SI{3.5}{\MeV}$, respectively, see
Tab.~\ref{tab:tc-ontheway}. It is, however, interesting to note that the linear
extrapolation of our results towards the chiral limit, shown in right diagram of
Fig.~\ref{fig:msp-susc}, underestimates the chiral transition temperature
calculated explicitly in the chiral light-quark limit by about two MeV. The
linear extrapolation results in $\SI{145.4}{\MeV}$, whereas we find the value
\begin{equation}
    \Tc(\massl=0) = \SI{146.7}{\MeV}
    \,,
\end{equation}
which is about $\SIrange{5}{5.5}{\MeV}$ larger than the FRG and FRG--DSE
results, respectively, and more than ten MeV larger than the extrapolated
lattice value $\Tc^{\text{HotQCD}} = 132^{+3}_{-6}~\si{\MeV}$. We
attribute this to our vertex construction, which contains a strength parameter
$d_1$ which is fixed at the physical point and not changed with quark mass. We
therefore slightly overestimate the interaction strength in the chiral limit
leading to slightly too large transition temperatures. This will be discussed
again also in the next section.

\subsection{\label{subsec:left-edge}%
    Left edge of the Columbia plot
}

Next, we turn our analysis to the left edge of the Columbia plot. To this end,
we display the temperature dependence of the quark condensate for chiral light
quarks and a set of six selected strange-quark masses between $\masss \in [0,
\infty)$ in the left diagram of Fig.~\ref{fig:mu0-Tc-vs-ms}. One can
immediately notice that for all investigated strange-quark masses we observe a
second-order phase transition. That is, the quark condensate continuously
changes from a nonzero value to zero with increasing temperature with no
(apparent) jumps. As we will see in Sec.~\ref{subsec:scaling}, it is indeed a
genuine second-order transition since the condensate exhibits a scaling behavior
in the vicinity of the respective critical temperatures. We emphasize that this
also holds true for the $\Nf = 3$ corner where we consequently find no
first-order region at all. In general, the condensate is smaller for smaller
strange-quark masses at all temperatures. The only exception occurs close to the
three-flavor limit at around $\masss \sim \SI{e-9}{\MeV}$ where we do find
a sudden and small increase in the condensate for all temperatures which then
remains constant until $\masss \rightarrow 0$. In the left diagram of
Fig.~\ref{fig:mu0-Tc-vs-ms}, this increase is visible when comparing the $\masss
= \SI{e-3}{\MeV}$ result with the the one at $\masss = \SI{0}{\MeV}$. Since we
neither found a technical nor a physical reason for this glitch we attribute it
(for the moment) to a pure numerical artefact of the three-flavor limit.

The dependence of the critical temperature on the strange-quark mass is
illustrated in the right diagram of Fig.~\ref{fig:mu0-Tc-vs-ms}. Qualitatively,
we find that $\Tc$ varies very little for very small and very high strange-quark
masses but increases monotonously in the range $\masss
=\qtyrange[print-unity-mantissa=false]{1}{e4}{\MeV}$.

In Tab.~\ref{tab:tc-ms}, we compare our findings of the critical temperature
quantitatively to the most-recent lattice results available for zero, physical
and infinite strange-quark masses. We observe that our values for $\Tc$ are
consistently larger than the ones found on the lattice, with the smallest
difference at the physical strange-quark mass. Our explanation for this
discrepancy is based on the discussion above: we fix the interaction strength
$d_1$ for the non-hadronic part of our quark--gluon interaction, Eq.~(\ref{d1}),
at the physical point and do not take into account any changes of the vertex
strength with variation of the quark masses. Presumably, this leads to the small
discrepancy in transition temperature in the light chiral limit with physical
strange quarks already discussed above and larger discrepancies in the chiral
corners of the Columbia plot. We have explicitly checked what happens in the
$\Nf=3$ limit, when we adapt the vertex strength. With $d_1 =
\SI{7.13}{\GeV^{2}}$ we reproduce the transition temperature of
Ref.~\cite{Dini:2021hug} while the transition is still second order. Thus, the
value of $d_1$ (at least within the range studied here) had no material
influence on the order of the transition.

Finally, we investigated whether the fate of the $\UA(1)$ symmetry has any
influence on the order of the transition. On the complexity level of the present
truncation, an anomalously broken $\UA(1)$ at the critical temperature results
in a massive $\eta'$ meson, whereas a restored $\UA(1)$ renders the $\eta'$
meson massless. So far, we assumed the first case and neglected the $\eta'$
meson in the backreaction diagrams completely together with all other mesons
that remain massive in the chiral $\Nf=3$ limit. In order to gauge the influence
of the $\eta'$ on the order of the transition in this limit, we repeated our
calculation with the massive $\eta'$ and, even more importantly, with a massless
$\eta'$ explicitly present in the loops. As a result, we find only a very small
reduction of the transition temperature of about $\SI{0.3}{\MeV}$ when including
a massive $\eta'$ with no changes in the second-order nature of the transition.
This result confirms our notion that additional massive mesons barely have any
influence on our results and their omission, therefore, is a good approximation.
Including a massless $\eta'$ reduces the transition temperature further by about
$\SI{1.5}{\MeV}$ but again does not change the second-order nature of the phase
transition. We therefore conclude that within the framework presented here the
fate of the anomalously broken axial $\UA(1)$ symmetry with temperature has no
material effect on the order of the chiral phase transition.

\begin{table}
	\centering
	\def\arraystretch{1.5}
	\begin{tabular}{|c|c|c|c|c|}
		\hline
		$\masss$                    &         & $0$                                & $\masss^{p}$                         & $\infty$ \\
		\hline\cline{1-5}
		\multirow{2}{*}{$\Tc \ [\SI{}{MeV}]$}
		& DSE     & $133.4$                            & $146.7$                              & $204.2$ \\
		\cline{2-5}
		& Lattice & $98_{-6}^{+3}$ \cite{Dini:2021hug} & $132_{-6}^{+3}$\cite{HotQCD:2019xnw} & $174 \pm 3 \pm 6$ \cite{Bornyakov:2009qh} \\
		\hline
	\end{tabular}
	\caption{\label{tab:tc-ms}%
		Comparison of critical temperatures for different strange-quark masses
		between our DSE findings and lattice results. Adapting our vertex strength parameter
		to match the lattice critical temperature in the $m_s=0$-limit does not change the
		second-order nature of the chiral phase transition.
	}
\end{table}

\subsection{\label{subsec:scaling}%
    Scaling behavior
}

\begin{figure*}
	\centering
	\includegraphics[scale=1.0]{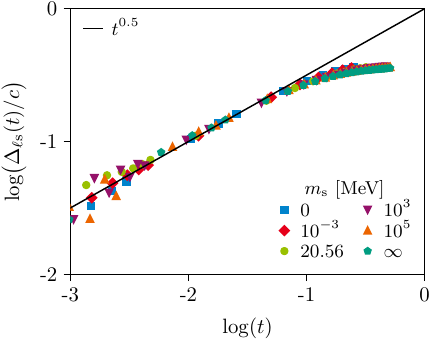}
	\hfil
	\includegraphics[scale=1.0]{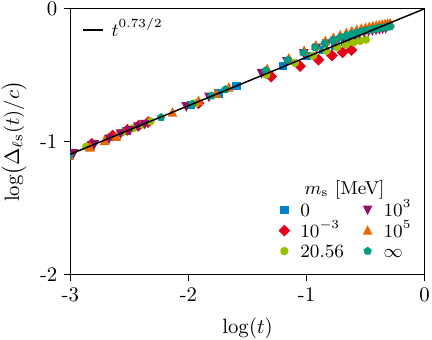}
	\caption{\label{fig:mu0-scale}%
		\emph{Left}: Scaling behavior of the regularized quark condensate as a
		function of the reduced temperature for different strange-quark masses
		in the up quark chiral limit without scaling decay constants. %
		\emph{Right}: Same scaling behavior with scaling decay constants (see
		main text for discussion).%
	}
\end{figure*}

In the vicinity of a second-order phase transition the order parameter, obeys a
power law with respect to some (universal) reduced quantity. For the regularized
condensate, we expect the following behavior:
\begin{equation}\label{eq:scaling-relation}
    \Delta_{\ell\ups}(T)
    \sim
    c \,
    t^{\beta}
    \,,
    \quad
    \text{where}
    \quad
    t
    =
    \frac{\Tc - T}{\Tc}
\end{equation}
labels the reduced temperature, $\beta$ indicates the critical exponent
depending on the underlying universality class and $c$ denotes a non-universal
constant. The scaling properties of the quark DSE and, related, that of the
condensate have been studied in Ref.~\cite{Fischer:2011pk} in the chiral $\Nf=2$
limit. Here we expect a second-order phase transition in the $\OO(4)$
universality class of the Heisenberg antiferromagnet, see, e.g.,
Ref.~\cite{Pisarski:1983ms}. Indeed, it has been shown in
Ref.~\cite{Fischer:2011pk} that the correct scaling is obtained if (and only if)
the scaling of the temporal meson decay constants is taken into account.
Therefore, to obtain self-consistent scaling from the quark DSE, one would need
to explicitly include the BSE for the relevant long-range degrees of freedom,
the pion and the sigma, as well as the corresponding equations for their decay
constants in a self-consistent manner. This is beyond the present truncation and
would require extensive additional work. A shortcut, also used in
Ref.~\cite{Fischer:2011pk}, is to assume the critical scaling of the decay
constants and only check for consistency in the quark DSE using an appropriate
ansatz for this scaling. Indeed, one then finds that the diagrams with
long-range massless meson degrees of freedom dominate over the gluon ones, i.e.,
universality sets in and scaling is observed. On the other hand, any setup
without scaling decay constants delivers the usual mean-field scaling observed
in rainbow--ladder type truncations already very early and reviewed in
Ref.~\cite{Roberts:2000aa}.

For completeness, we have checked both, mean field scaling without and
$\OO(4)$ scaling with proper scaling decay constants. For the former we use the
Pagels--Stokar approximation for the decay constants discussed in
Eq.~(\ref{eq:pagels-stokar-actual}), for the latter we use the following
ansatz\footnote{%
    Indeed, this is precisely the behavior we also find when calculating the
    decay constants using the Pagels--Stokar formula except with a mean-field
    critical exponent of $\beta = 0.5$.%
} %
\begin{align}\label{eq:scale}
    \tilde{f}_{\ell \ell}^{Y}(T)
    &=
    f_{\ell \ell}^{Y}(T_{0})
    \biggl(
    \frac{\Tc - T}{\Tc - T_{0}}
    \biggr)^{\beta}
    \,,
    \\
    \tilde{f}_{\ell \ups}^{\textrm{ps}}(T)
    &=
    f_{\ell \ups}^{\textrm{ps}}(T_{0})
    \biggl(
    \frac{\Tc - T}{\Tc - T_{0}}
    \biggr)^{\beta / 2}
    \,,
    \\
    \tilde{f}^{Y}_{xy}(T)
    &=
    f^{Y}_{xy}(T_{0})
    \,.
\end{align}
Here, we use the decay constants from the Pagels--Stokar setup at some
temperature $T_{0} = \SI{100}{\MeV}$ as input, while the critical temperatures
$\Tc$ are the ones from Fig.~\ref{fig:mu0-Tc-vs-ms}. The scaling law for
$f_{\ell \ell}^{Y}$ is taken from Ref.~\cite{Fischer:2011pk}, whereas the
scaling law for $f_{\ell \ups}^{\textrm{ps}}$, valid for $m_s \ne 0$ is obtained
from an analogous scaling analysis as in Ref.~\cite{Fischer:2011pk} for the kaon
diagrams. All other decay constants do not exhibit any critical scaling. In the
limit of $m_s \rightarrow 0$, we just assume the same critical scaling for
$f_{\ell \ell}^{Y}$, $f_{\ell \ups}^{\textrm{ps}}$ and $f_{\ups
\ups}^{\textrm{ps}}$. This is, however, for simplicity only since the
universality class in this limit is not known.

In Fig.~\ref{fig:mu0-scale}, we display our results. In the left diagram, we
show the (logarithm of the) regularized condensates in the chiral light-quark
limit and for the same strange-quark masses as in Fig.~\ref{fig:mu0-Tc-vs-ms} as
functions of the (logarithm of the) reduced temperature $t$. For the sake of
comparability, we fit each dataset to the relation in
Eq.~\eqref{eq:scaling-relation} and divide by their respective constant $c$. As
can be seen, all curves collapse nicely onto each other and align along the line
$t^{0.5}$ for $\log(t) \leq -1.5$. The spread of data points below $\log(t) \leq
-2$ is entirely due to our numerical uncertainty of $\Delta \Tc =
\SI{0.1}{\MeV}$ in the determination of $\Tc$. We observe the expected
mean-field scaling behavior for all investigated strange-quark masses.

In the right diagram of Fig.~\ref{fig:mu0-scale}, we show corresponding results
under the assumption that the decay constants scale according to
Eq.~(\ref{eq:scale}) with the correct $\OO(4)$ critical exponents of QCD, where
$\beta = 0.73 / 2$. As expected, the scaling behavior of the decay constants
induces a consistent corresponding scaling of the order parameter. Furthermore,
with induced scaling, the collapse of all datasets for $\log(t) \leq -1.5$ is
almost perfect and the spread for low $\log(t)$ vanishes completely. This is due
to the fact that the additional appearance of $T_c$ in the scaling ansatz
stabilizes the numerics considerably.

\begin{figure}
    \centering
    \includegraphics[scale=1.0]{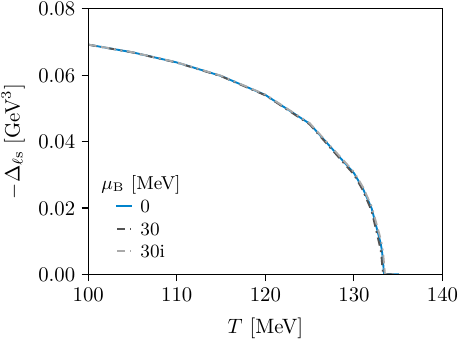}
    \\[0.5em]
    \includegraphics[scale=1.0]{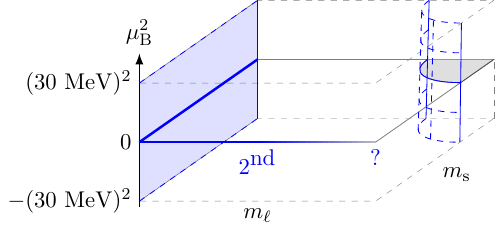}
    \caption{\label{fig:mu0-imuu}%
        Regularized quark condensate for (small) chemical potentials in the up-
        and strange-quark chiral limits. %
        \emph{Top}: Regularized quark condensate as a function of temperature. %
        \emph{Bottom}:
        Illustration of the three-dimensional Columbia plot we find. %
    }
\end{figure}

Of course, since $\beta = 0.73 / 2$ is an input, this setup reveals nothing
about the universality class in the chiral three-flavor limit. In order to study
this issue from DSEs, as discussed above, one needs to solve the corresponding
Bethe--Salpeter equations and the defining equations for the decay constants
without further approximations. This is possible, in principle, and should be
attempted in future work. In practice, it may however be more straightforward to
perform this calculation in the framework of the functional renormalization
group, where scaling properties are more directly approachable
\cite{Berges:2000ew,Schaefer:2004en,Schaefer:2006ds,Schaefer:2008hk,Resch:2017vjs,Rennecke:2016tkm,Braun:2020ada}.

\subsection{\label{subsec:chempot}%
    Three-dimensional Columbia plot
}

Finally, we would like to explore the fate of the second-order phase transition
along the left-hand side of the Columbia plot when we switch on chemical
potential. Is there a second-order critical sheet? And if yes, does it bend at
some point towards nonzero quark masses and is it connected to the CEP that we
find at physical quark masses
\cite{Fischer:2012vc,Isserstedt:2019pgx,Gunkel:2021oya}?

Unfortunately, these questions are difficult to study. Our current approximation
for the meson Bethe--Salpeter amplitudes, Eq.~(\ref{eq:goldberger-treiman}), is
known to be accurate at vanishing chemical potential. From the explicit
calculation in Ref.\cite{Gunkel:2019xnh,Gunkel:2020wcl}, however, it is known
that the amplitudes are modified substantially at large chemical potential. We
can therefore only trust the approximation Eq.~(\ref{eq:goldberger-treiman}) at
small chemical potentials.

We therefore restrict ourselves to real and imaginary baryon chemical potentials
of $|\muB| =\SI{30}{\MeV}$. The corresponding results are shown in
Fig.~\ref{fig:mu0-imuu} together with results for $\muB = 0$ as a comparison.
The calculations have been performed for $\masss = 0, \masss^{p}, \infty$. In
the top panel, we display the condensate as a function of temperature in the
$\Nf=3$ chiral limit, i.e., $\massl=\masss = 0$. We find no significant changes
within this range of chemical potential. Similar results are obtained for all
investigated strange-quark masses in $\masss = 0, \masss^{p}, \infty$. In total,
we find little quantitative and no qualitative difference between the results
for vanishing and small chemical potentials. That is, one can still observe a
second-order phase transition with identical scaling behavior and an almost
unchanged critical temperature. We therefore arrive at the slice of the
three-dimensional Columbia plot shown in the bottom panel of
Fig.~\ref{fig:mu0-imuu}. For zero chemical potential, this ties in with the
lattice results of Ref.~\cite{Cuteri:2021ikv,Dini:2021hug} and for imaginary
chemical potential with Ref.~\cite{DAmbrosio:2022kig}. It also agrees with one
of the scenarios displayed in the FRG approach in Ref.~\cite{Resch:2017vjs}
(their right diagram of Fig. 3), but disagrees with the other scenarios they
give. This needs to be reexamined in some detail. In any case, the analyticity
of the second-order transition plane from small imaginary to small real chemical
potential visible in Fig.~\ref{fig:mu0-imuu} is, to our knowledge, shown for the
first time.

Last, we perform a first exploration of the `bottom' line of the Columbia plot,
i.e., the zero chemical-potential line with a massless strange quark and
up/down-quark masses in the range $\massl = 0,..., \massl^{p},..., \infty$.
Then, the strange-quark condensate becomes the corresponding order parameter for
the chiral transition at finite temperature, so we include one massless
Goldstone boson due to the dynamical breaking of a $\UA(1)$ subgroup of the
flavor $\SUA(3)$ and we expect the isoscalar scalar $f_0$ with $\ups\bar{\ups}$-
content to be the only additional massless mode at the critical temperature.
Furthermore, we assume a restored anomaly at the critical temperature. The
corresponding meson masses, flavor coefficients and decay constants are detailed
in Appendix~\ref{appendix}. For this setup, we indeed find again a second-order
phase transition, also indicated in our three-dimensional Columbia plot in the
bottom panel of Fig.~\ref{fig:mu0-imuu}. This second-order transition persists
to large but not infinite up/down-quark masses. At some large up/down-quark
mass, $\massl > \SI{100}{\GeV}$, our calculation breaks down indicating that we
are approaching the one-flavor limit of QCD with different symmetries. The
detailed study of this limit is non-trivial and deferred to future work.

\section{\label{sec:summary}%
    Summary and conclusions
}

In this work, we studied the order of the phase transition in the light chiral
limit of massless up/down quarks as a function of the mass of the strange quark
and at zero and small values for the baryon chemical potential. Using a
truncation of Dyson--Schwinger equations that takes into account microscopic
degrees of freedom as well as potential long-range correlations with the quantum
numbers of pseudoscalar and scalar mesons, we obtain a chiral crossover as long
as the light-quark masses remain nonzero, but a second-order phase transition in
the light chiral limit. This behavior persists along the left-hand side of the
Columbia plot, i.e., for all strange-quark masses in $0 \le m_s \le \infty$ and
also for (small) imaginary and real chemical potential. It persists regardless
whether we include a massive $\eta'$ meson (in case the axial $\UA(1)$ remains
anomalously broken at $T_c$) or a massless $\eta'$ meson (in case the axial
$\UA(1)$ is restored at $T_c$). Our findings do not support the long-standing
notion of a chiral first-order $\Nf=3$ corner in the Columbia plot
\cite{Pisarski:1983ms}, but agree with recent findings from lattice QCD
\cite{Cuteri:2021ikv,Dini:2021hug} and notions from effective models
\cite{Fejos:2022mso}.


\begin{acknowledgments}
We thank Fabian Rennecke, Bernd-Jochen Schaefer, Jan Pawlowski, Owe Philipsen
and Lorenz von Smekal for very fruitful discussions and Bernd-Jochen Schaefer
and Owe Philipsen for a critical reading of the manuscript. We thank Owe
Philipsen for suggesting the extension to (small) chemical potential performed
in this work. We furthermore thank Philipp Isserstedt for cross-checks of our
code at an early stage of this work. This work has been supported by the
Helmholtz Graduate School for Hadron and Ion Research (HGS-HIRe) for FAIR, the
GSI Helmholtzzentrum f\"{u}r Schwerionenforschung, and the Deutsche
Forschungsgemeinschaft (DFG, German Research Foundation) through the
Collaborative Research Center TransRegio CRC-TR 211 ``Strong-interaction matter
under extreme conditions'' and the individual grant FI 970/16-1. This work is
furthermore part of a project that has received funding from the European
Union’s Horizon 2020 research and innovation programme under grant agreement
STRONG – 2020 - No 824093. Feynman diagrams were drawn with
\textsc{TikZ-Feynman} \cite{Ellis:2016jkw}.
\end{acknowledgments}


\appendix

\section{\label{appendix}%
    Meson masses and flavor coefficients of the $\Nf=1+2$ case
}

\begin{table}
    \centering
    \setlength{\tabcolsep}{4pt}
    \def\arraystretch{1.25}\hspace*{-3mm}
    \begin{minipage}[t]{0.225\textwidth}
        $\Nf = 1 + 2$:\\[0.25em]
        \begin{tabular}{|c|c||c|c|c|}
            \hline
            $f$ & $X$ & $F_{X}^{f}$ & $S_{X}^{f}$ & $f_{X}^{f}$ \\
            \hline
            $\ell$ & $\pion$   & $3 / 2$ & $S_{\ell}$ & $f_{\ell\ell}^{\Pps}$ \\
            $\ell$ & $\sigma$  & $1/2$ & $S_{\ell}$ & $f_{\ell\ell}^{\Psc}$ \\
            $\ell$ & $\kaon$   & $1$     & $S_{\ups}$ & $f_{\ell\ups}^{\Pps}$ \\
            $\ell$ & $\etalmes$ & $1 /2$ & $S_{\ell}$ & $f_{\ell\ell}^{\Pps}$ \\
            $\ups$ & $\kaon$   & $2$     & $S_{\ell}$ & $f_{\ups\ell}^{\Pps}$ \\
            $\ups$ & $\etasmes$& $1$ & $S_{\ups}$ & $f_{\ups\ups}^{\Pps}$ \\
            $\ups$ & $\fmes$   & $1 / 2$ & $S_{\ups}$ & $f_{\ups\ups}^{\Psc}$ \\
            \hline
        \end{tabular}
    \end{minipage}
    \begin{minipage}[t]{0.225\textwidth}
        $\Nf = 1 + 2 \to 3$:\\[0.25em]
        \begin{tabular}{|c|c||c|c|c|}
            \hline
            $f$ & $X$ & $F_{X}^{f}$ & $S_{X}^{f}$ & $f_{X}^{f}$ \\
            \hline
            $\ell$ & $\pion, \kaon, \etalmes$  & $3$ & $S_{\ell}$ & $f_{\ell\ell}^{\Pps}$ \\
            $\ell$ & $\sigma$ & $1 / 2$ & $S_{\ell}$ & $f_{\ell\ell}^{\Psc}$ \\
            \hline
        \end{tabular}
    \end{minipage}
    \caption{\label{tab:meson-info-nf21}%
        Information of multiplicities, internal quark propagators and decay
        constants for all considered meson backcoupling diagrams in the $\Nf =
        1 + 2$ setup.%
    }
\end{table}

For the $\Nf = 1 + 2$ setup of Sec.~\ref{subsec:chempot} with a chiral strange
quark and a varying up/down-quark mass, we work with two assumptions. First, we
assume that the axial anomaly is restored at the chiral transition temperature
such that no anomalous mass contributions arise. Second, under this assumption,
it is natural to assume that mixing between the isoscalar, pseudoscalar octet
and singlet states results in a pure $\ups \bar{\ups}$ massless Goldstone boson
and a massive meson with up/down-quark content, i.e., $(\etames, \etapmes) \to
(\etalmes, \etasmes)$. The resulting multiplicities are displayed in
Tab.~\ref{tab:meson-info-nf21} (analogously to Tab.~\ref{tab:meson-info}). The
limit $N_f=1+2 \rightarrow 3$ is consistent with the corresponding limit
$N_f=2+1 \rightarrow 3$ in Tab.~\ref{tab:meson-info-nf21} under inclusion of the
$\eta_0$. In the scalar meson sector, a massless isoscalar $\ups \bar{\ups}$
(which we call $f_0$) at the chiral transition temperature reflects the fact
that the strange-quark condensate is the appropriate order parameter. The kaons
are massive away from the $N_f=3$ limit, similar to the $N_f=2+1$ case, but this
time due to the non-chiral up/down quarks. In total, we reuse the
parametrizations of Eq.~\eqref{masses}) but adjust the quark masses in the kaon
argument and set the $\fmes$ and $\etasmes$ masses to zero:
\begin{align}
\begin{split}
    m_{\pion}
    &=
    \SI{156.525}{\MeV^{1/2}}
    \cdot
    \sqrt{\massl}
    \,,
    \quad
    m_{\sigma}
    =
    2 m_{\pion}
    \,,
    \\
    m_{\kaon}
    &=
    \SI{74.2}{\MeV^{1/2}}
    \cdot
    \sqrt{\massl}
    +
    1.54
    \cdot
    \massl
    \,,
    \\
    m_{\etalmes}
    &=
    2 m_{\kaon}
    \,,
    \quad
    m_{\etasmes}
    =
    m_{\fmes}
    =
    0
    \,.
\end{split}
\end{align}

\bibliography{ColumbiaMesonBCBibliography}

\begin{thebibliography}{104}%
\makeatletter
\providecommand \@ifxundefined [1]{%
 \@ifx{#1\undefined}
}%
\providecommand \@ifnum [1]{%
 \ifnum #1\expandafter \@firstoftwo
 \else \expandafter \@secondoftwo
 \fi
}%
\providecommand \@ifx [1]{%
 \ifx #1\expandafter \@firstoftwo
 \else \expandafter \@secondoftwo
 \fi
}%
\providecommand \natexlab [1]{#1}%
\providecommand \enquote  [1]{``#1''}%
\providecommand \bibnamefont  [1]{#1}%
\providecommand \bibfnamefont [1]{#1}%
\providecommand \citenamefont [1]{#1}%
\providecommand \href@noop [0]{\@secondoftwo}%
\providecommand \href [0]{\begingroup \@sanitize@url \@href}%
\providecommand \@href[1]{\@@startlink{#1}\@@href}%
\providecommand \@@href[1]{\endgroup#1\@@endlink}%
\providecommand \@sanitize@url [0]{\catcode `\\12\catcode `\$12\catcode
  `\&12\catcode `\#12\catcode `\^12\catcode `\_12\catcode `\%12\relax}%
\providecommand \@@startlink[1]{}%
\providecommand \@@endlink[0]{}%
\providecommand \url  [0]{\begingroup\@sanitize@url \@url }%
\providecommand \@url [1]{\endgroup\@href {#1}{\urlprefix }}%
\providecommand \urlprefix  [0]{URL }%
\providecommand \Eprint [0]{\href }%
\providecommand \doibase [0]{http://dx.doi.org/}%
\providecommand \selectlanguage [0]{\@gobble}%
\providecommand \bibinfo  [0]{\@secondoftwo}%
\providecommand \bibfield  [0]{\@secondoftwo}%
\providecommand \translation [1]{[#1]}%
\providecommand \BibitemOpen [0]{}%
\providecommand \bibitemStop [0]{}%
\providecommand \bibitemNoStop [0]{.\EOS\space}%
\providecommand \EOS [0]{\spacefactor3000\relax}%
\providecommand \BibitemShut  [1]{\csname bibitem#1\endcsname}%
\let\auto@bib@innerbib\@empty
\bibitem [{\citenamefont {Bzdak}\ \emph {et~al.}(2020)\citenamefont {Bzdak},
  \citenamefont {Esumi}, \citenamefont {Koch}, \citenamefont {Liao},
  \citenamefont {Stephanov},\ and\ \citenamefont {Xu}}]{Bzdak:2019pkr}%
  \BibitemOpen
  \bibfield  {author} {\bibinfo {author} {\bibfnamefont {A.}~\bibnamefont
  {Bzdak}}, \bibinfo {author} {\bibfnamefont {S.}~\bibnamefont {Esumi}},
  \bibinfo {author} {\bibfnamefont {V.}~\bibnamefont {Koch}}, \bibinfo {author}
  {\bibfnamefont {J.}~\bibnamefont {Liao}}, \bibinfo {author} {\bibfnamefont
  {M.}~\bibnamefont {Stephanov}}, \ and\ \bibinfo {author} {\bibfnamefont
  {N.}~\bibnamefont {Xu}},\ }\href {\doibase 10.1016/j.physrep.2020.01.005}
  {\bibfield  {journal} {\bibinfo  {journal} {Phys. Rept.}\ }\textbf {\bibinfo
  {volume} {853}},\ \bibinfo {pages} {1} (\bibinfo {year} {2020})},\ \Eprint
  {http://arxiv.org/abs/1906.00936} {arXiv:1906.00936 [nucl-th]} \BibitemShut
  {NoStop}%
\bibitem [{\citenamefont {Friman}\ \emph {et~al.}(2011)\citenamefont {Friman},
  \citenamefont {Hohne}, \citenamefont {Knoll}, \citenamefont {Leupold},
  \citenamefont {Randrup}, \citenamefont {Rapp},\ and\ \citenamefont
  {Senger}}]{Friman:2011zz}%
  \BibitemOpen
  \bibinfo {editor} {\bibfnamefont {B.}~\bibnamefont {Friman}}, \bibinfo
  {editor} {\bibfnamefont {C.}~\bibnamefont {Hohne}}, \bibinfo {editor}
  {\bibfnamefont {J.}~\bibnamefont {Knoll}}, \bibinfo {editor} {\bibfnamefont
  {S.}~\bibnamefont {Leupold}}, \bibinfo {editor} {\bibfnamefont
  {J.}~\bibnamefont {Randrup}}, \bibinfo {editor} {\bibfnamefont
  {R.}~\bibnamefont {Rapp}}, \ and\ \bibinfo {editor} {\bibfnamefont
  {P.}~\bibnamefont {Senger}},\ eds.,\ \href {\doibase
  10.1007/978-3-642-13293-3} {\emph {\bibinfo {title} {The CBM physics book:
  Compressed baryonic matter in laboratory experiments}}},\ Vol.\ \bibinfo
  {volume} {814}\ (\bibinfo {year} {2011})\BibitemShut {NoStop}%
\bibitem [{\citenamefont {Salabura}\ and\ \citenamefont
  {Stroth}(2021)}]{Salabura:2020tou}%
  \BibitemOpen
  \bibfield  {author} {\bibinfo {author} {\bibfnamefont {P.}~\bibnamefont
  {Salabura}}\ and\ \bibinfo {author} {\bibfnamefont {J.}~\bibnamefont
  {Stroth}},\ }\href {\doibase 10.1016/j.ppnp.2021.103869} {\bibfield
  {journal} {\bibinfo  {journal} {Prog. Part. Nucl. Phys.}\ }\textbf {\bibinfo
  {volume} {120}},\ \bibinfo {pages} {103869} (\bibinfo {year} {2021})},\
  \Eprint {http://arxiv.org/abs/2005.14589} {arXiv:2005.14589 [nucl-ex]}
  \BibitemShut {NoStop}%
\bibitem [{\citenamefont {Almaalol}\ \emph {et~al.}(2022)\citenamefont
  {Almaalol} \emph {et~al.}}]{Almaalol:2022xwv}%
  \BibitemOpen
  \bibfield  {author} {\bibinfo {author} {\bibfnamefont {D.}~\bibnamefont
  {Almaalol}} \emph {et~al.},\ }\href@noop {} {\  (\bibinfo {year} {2022})},\
  \Eprint {http://arxiv.org/abs/2209.05009} {arXiv:2209.05009 [nucl-ex]}
  \BibitemShut {NoStop}%
\bibitem [{\citenamefont {Braun-Munzinger}\ \emph {et~al.}(2016)\citenamefont
  {Braun-Munzinger}, \citenamefont {Koch}, \citenamefont {Sch\"afer},\ and\
  \citenamefont {Stachel}}]{Braun-Munzinger:2015hba}%
  \BibitemOpen
  \bibfield  {author} {\bibinfo {author} {\bibfnamefont {P.}~\bibnamefont
  {Braun-Munzinger}}, \bibinfo {author} {\bibfnamefont {V.}~\bibnamefont
  {Koch}}, \bibinfo {author} {\bibfnamefont {T.}~\bibnamefont {Sch\"afer}}, \
  and\ \bibinfo {author} {\bibfnamefont {J.}~\bibnamefont {Stachel}},\ }\href
  {\doibase 10.1016/j.physrep.2015.12.003} {\bibfield  {journal} {\bibinfo
  {journal} {Phys. Rept.}\ }\textbf {\bibinfo {volume} {621}},\ \bibinfo
  {pages} {76} (\bibinfo {year} {2016})},\ \Eprint
  {http://arxiv.org/abs/1510.00442} {arXiv:1510.00442 [nucl-th]} \BibitemShut
  {NoStop}%
\bibitem [{\citenamefont {Luo}\ and\ \citenamefont {Xu}(2017)}]{Luo:2017faz}%
  \BibitemOpen
  \bibfield  {author} {\bibinfo {author} {\bibfnamefont {X.}~\bibnamefont
  {Luo}}\ and\ \bibinfo {author} {\bibfnamefont {N.}~\bibnamefont {Xu}},\
  }\href {\doibase 10.1007/s41365-017-0257-0} {\bibfield  {journal} {\bibinfo
  {journal} {Nucl. Sci. Tech.}\ }\textbf {\bibinfo {volume} {28}},\ \bibinfo
  {pages} {112} (\bibinfo {year} {2017})},\ \Eprint
  {http://arxiv.org/abs/1701.02105} {arXiv:1701.02105 [nucl-ex]} \BibitemShut
  {NoStop}%
\bibitem [{\citenamefont {Borsanyi}\ \emph
  {et~al.}(2010{\natexlab{a}})\citenamefont {Borsanyi}, \citenamefont {Fodor},
  \citenamefont {Hoelbling}, \citenamefont {Katz}, \citenamefont {Krieg},
  \citenamefont {Ratti},\ and\ \citenamefont {Szabo}}]{Borsanyi:2010bp}%
  \BibitemOpen
  \bibfield  {author} {\bibinfo {author} {\bibfnamefont {S.}~\bibnamefont
  {Borsanyi}}, \bibinfo {author} {\bibfnamefont {Z.}~\bibnamefont {Fodor}},
  \bibinfo {author} {\bibfnamefont {C.}~\bibnamefont {Hoelbling}}, \bibinfo
  {author} {\bibfnamefont {S.~D.}\ \bibnamefont {Katz}}, \bibinfo {author}
  {\bibfnamefont {S.}~\bibnamefont {Krieg}}, \bibinfo {author} {\bibfnamefont
  {C.}~\bibnamefont {Ratti}}, \ and\ \bibinfo {author} {\bibfnamefont {K.~K.}\
  \bibnamefont {Szabo}} (\bibinfo {collaboration} {Wuppertal-Budapest}),\
  }\href {\doibase 10.1007/JHEP09(2010)073} {\bibfield  {journal} {\bibinfo
  {journal} {JHEP}\ }\textbf {\bibinfo {volume} {09}},\ \bibinfo {pages} {073}
  (\bibinfo {year} {2010}{\natexlab{a}})},\ \Eprint
  {http://arxiv.org/abs/1005.3508} {arXiv:1005.3508 [hep-lat]} \BibitemShut
  {NoStop}%
\bibitem [{\citenamefont {Bazavov}\ \emph
  {et~al.}(2012{\natexlab{a}})\citenamefont {Bazavov} \emph
  {et~al.}}]{Bazavov:2011nk}%
  \BibitemOpen
  \bibfield  {author} {\bibinfo {author} {\bibfnamefont {A.}~\bibnamefont
  {Bazavov}} \emph {et~al.},\ }\href {\doibase 10.1103/PhysRevD.85.054503}
  {\bibfield  {journal} {\bibinfo  {journal} {Phys. Rev. D}\ }\textbf {\bibinfo
  {volume} {85}},\ \bibinfo {pages} {054503} (\bibinfo {year}
  {2012}{\natexlab{a}})},\ \Eprint {http://arxiv.org/abs/1111.1710}
  {arXiv:1111.1710 [hep-lat]} \BibitemShut {NoStop}%
\bibitem [{\citenamefont {Borsanyi}\ \emph
  {et~al.}(2010{\natexlab{b}})\citenamefont {Borsanyi}, \citenamefont
  {Endrodi}, \citenamefont {Fodor}, \citenamefont {Jakovac}, \citenamefont
  {Katz}, \citenamefont {Krieg}, \citenamefont {Ratti},\ and\ \citenamefont
  {Szabo}}]{Borsanyi:2010cj}%
  \BibitemOpen
  \bibfield  {author} {\bibinfo {author} {\bibfnamefont {S.}~\bibnamefont
  {Borsanyi}}, \bibinfo {author} {\bibfnamefont {G.}~\bibnamefont {Endrodi}},
  \bibinfo {author} {\bibfnamefont {Z.}~\bibnamefont {Fodor}}, \bibinfo
  {author} {\bibfnamefont {A.}~\bibnamefont {Jakovac}}, \bibinfo {author}
  {\bibfnamefont {S.~D.}\ \bibnamefont {Katz}}, \bibinfo {author}
  {\bibfnamefont {S.}~\bibnamefont {Krieg}}, \bibinfo {author} {\bibfnamefont
  {C.}~\bibnamefont {Ratti}}, \ and\ \bibinfo {author} {\bibfnamefont {K.~K.}\
  \bibnamefont {Szabo}},\ }\href {\doibase 10.1007/JHEP11(2010)077} {\bibfield
  {journal} {\bibinfo  {journal} {JHEP}\ }\textbf {\bibinfo {volume} {11}},\
  \bibinfo {pages} {077} (\bibinfo {year} {2010}{\natexlab{b}})},\ \Eprint
  {http://arxiv.org/abs/1007.2580} {arXiv:1007.2580 [hep-lat]} \BibitemShut
  {NoStop}%
\bibitem [{\citenamefont {Borsanyi}\ \emph {et~al.}(2014)\citenamefont
  {Borsanyi}, \citenamefont {Fodor}, \citenamefont {Hoelbling}, \citenamefont
  {Katz}, \citenamefont {Krieg},\ and\ \citenamefont
  {Szabo}}]{Borsanyi:2013bia}%
  \BibitemOpen
  \bibfield  {author} {\bibinfo {author} {\bibfnamefont {S.}~\bibnamefont
  {Borsanyi}}, \bibinfo {author} {\bibfnamefont {Z.}~\bibnamefont {Fodor}},
  \bibinfo {author} {\bibfnamefont {C.}~\bibnamefont {Hoelbling}}, \bibinfo
  {author} {\bibfnamefont {S.~D.}\ \bibnamefont {Katz}}, \bibinfo {author}
  {\bibfnamefont {S.}~\bibnamefont {Krieg}}, \ and\ \bibinfo {author}
  {\bibfnamefont {K.~K.}\ \bibnamefont {Szabo}},\ }\href {\doibase
  10.1016/j.physletb.2014.01.007} {\bibfield  {journal} {\bibinfo  {journal}
  {Phys. Lett. B}\ }\textbf {\bibinfo {volume} {730}},\ \bibinfo {pages} {99}
  (\bibinfo {year} {2014})},\ \Eprint {http://arxiv.org/abs/1309.5258}
  {arXiv:1309.5258 [hep-lat]} \BibitemShut {NoStop}%
\bibitem [{\citenamefont {Bazavov}\ \emph {et~al.}(2014)\citenamefont {Bazavov}
  \emph {et~al.}}]{HotQCD:2014kol}%
  \BibitemOpen
  \bibfield  {author} {\bibinfo {author} {\bibfnamefont {A.}~\bibnamefont
  {Bazavov}} \emph {et~al.} (\bibinfo {collaboration} {HotQCD}),\ }\href
  {\doibase 10.1103/PhysRevD.90.094503} {\bibfield  {journal} {\bibinfo
  {journal} {Phys. Rev. D}\ }\textbf {\bibinfo {volume} {90}},\ \bibinfo
  {pages} {094503} (\bibinfo {year} {2014})},\ \Eprint
  {http://arxiv.org/abs/1407.6387} {arXiv:1407.6387 [hep-lat]} \BibitemShut
  {NoStop}%
\bibitem [{\citenamefont {Ding}\ \emph {et~al.}(2015)\citenamefont {Ding},
  \citenamefont {Karsch},\ and\ \citenamefont {Mukherjee}}]{Ding:2015ona}%
  \BibitemOpen
  \bibfield  {author} {\bibinfo {author} {\bibfnamefont {H.-T.}\ \bibnamefont
  {Ding}}, \bibinfo {author} {\bibfnamefont {F.}~\bibnamefont {Karsch}}, \ and\
  \bibinfo {author} {\bibfnamefont {S.}~\bibnamefont {Mukherjee}},\ }\href
  {\doibase 10.1142/S0218301315300076} {\bibfield  {journal} {\bibinfo
  {journal} {Int. J. Mod. Phys. E}\ }\textbf {\bibinfo {volume} {24}},\
  \bibinfo {pages} {1530007} (\bibinfo {year} {2015})},\ \Eprint
  {http://arxiv.org/abs/1504.05274} {arXiv:1504.05274 [hep-lat]} \BibitemShut
  {NoStop}%
\bibitem [{\citenamefont {Bazavov}\ \emph
  {et~al.}(2017{\natexlab{a}})\citenamefont {Bazavov} \emph
  {et~al.}}]{Bazavov:2017dus}%
  \BibitemOpen
  \bibfield  {author} {\bibinfo {author} {\bibfnamefont {A.}~\bibnamefont
  {Bazavov}} \emph {et~al.},\ }\href {\doibase 10.1103/PhysRevD.95.054504}
  {\bibfield  {journal} {\bibinfo  {journal} {Phys. Rev. D}\ }\textbf {\bibinfo
  {volume} {95}},\ \bibinfo {pages} {054504} (\bibinfo {year}
  {2017}{\natexlab{a}})},\ \Eprint {http://arxiv.org/abs/1701.04325}
  {arXiv:1701.04325 [hep-lat]} \BibitemShut {NoStop}%
\bibitem [{\citenamefont {Bazavov}\ \emph {et~al.}(2019)\citenamefont {Bazavov}
  \emph {et~al.}}]{HotQCD:2018pds}%
  \BibitemOpen
  \bibfield  {author} {\bibinfo {author} {\bibfnamefont {A.}~\bibnamefont
  {Bazavov}} \emph {et~al.} (\bibinfo {collaboration} {HotQCD}),\ }\href
  {\doibase 10.1016/j.physletb.2019.05.013} {\bibfield  {journal} {\bibinfo
  {journal} {Phys. Lett. B}\ }\textbf {\bibinfo {volume} {795}},\ \bibinfo
  {pages} {15} (\bibinfo {year} {2019})},\ \Eprint
  {http://arxiv.org/abs/1812.08235} {arXiv:1812.08235 [hep-lat]} \BibitemShut
  {NoStop}%
\bibitem [{\citenamefont {Borsanyi}\ \emph {et~al.}(2020)\citenamefont
  {Borsanyi}, \citenamefont {Fodor}, \citenamefont {Guenther}, \citenamefont
  {Kara}, \citenamefont {Katz}, \citenamefont {Parotto}, \citenamefont
  {Pasztor}, \citenamefont {Ratti},\ and\ \citenamefont
  {Szabo}}]{Borsanyi:2020fev}%
  \BibitemOpen
  \bibfield  {author} {\bibinfo {author} {\bibfnamefont {S.}~\bibnamefont
  {Borsanyi}}, \bibinfo {author} {\bibfnamefont {Z.}~\bibnamefont {Fodor}},
  \bibinfo {author} {\bibfnamefont {J.~N.}\ \bibnamefont {Guenther}}, \bibinfo
  {author} {\bibfnamefont {R.}~\bibnamefont {Kara}}, \bibinfo {author}
  {\bibfnamefont {S.~D.}\ \bibnamefont {Katz}}, \bibinfo {author}
  {\bibfnamefont {P.}~\bibnamefont {Parotto}}, \bibinfo {author} {\bibfnamefont
  {A.}~\bibnamefont {Pasztor}}, \bibinfo {author} {\bibfnamefont
  {C.}~\bibnamefont {Ratti}}, \ and\ \bibinfo {author} {\bibfnamefont {K.~K.}\
  \bibnamefont {Szabo}},\ }\href {\doibase 10.1103/PhysRevLett.125.052001}
  {\bibfield  {journal} {\bibinfo  {journal} {Phys. Rev. Lett.}\ }\textbf
  {\bibinfo {volume} {125}},\ \bibinfo {pages} {052001} (\bibinfo {year}
  {2020})},\ \Eprint {http://arxiv.org/abs/2002.02821} {arXiv:2002.02821
  [hep-lat]} \BibitemShut {NoStop}%
\bibitem [{\citenamefont {Fischer}\ \emph {et~al.}(2014)\citenamefont
  {Fischer}, \citenamefont {Luecker},\ and\ \citenamefont
  {Welzbacher}}]{Fischer:2014ata}%
  \BibitemOpen
  \bibfield  {author} {\bibinfo {author} {\bibfnamefont {C.~S.}\ \bibnamefont
  {Fischer}}, \bibinfo {author} {\bibfnamefont {J.}~\bibnamefont {Luecker}}, \
  and\ \bibinfo {author} {\bibfnamefont {C.~A.}\ \bibnamefont {Welzbacher}},\
  }\href {\doibase 10.1103/PhysRevD.90.034022} {\bibfield  {journal} {\bibinfo
  {journal} {Phys. Rev. D}\ }\textbf {\bibinfo {volume} {90}},\ \bibinfo
  {pages} {034022} (\bibinfo {year} {2014})},\ \Eprint
  {http://arxiv.org/abs/1405.4762} {arXiv:1405.4762 [hep-ph]} \BibitemShut
  {NoStop}%
\bibitem [{\citenamefont {Isserstedt}\ \emph {et~al.}(2019)\citenamefont
  {Isserstedt}, \citenamefont {Buballa}, \citenamefont {Fischer},\ and\
  \citenamefont {Gunkel}}]{Isserstedt:2019pgx}%
  \BibitemOpen
  \bibfield  {author} {\bibinfo {author} {\bibfnamefont {P.}~\bibnamefont
  {Isserstedt}}, \bibinfo {author} {\bibfnamefont {M.}~\bibnamefont {Buballa}},
  \bibinfo {author} {\bibfnamefont {C.~S.}\ \bibnamefont {Fischer}}, \ and\
  \bibinfo {author} {\bibfnamefont {P.~J.}\ \bibnamefont {Gunkel}},\ }\href
  {\doibase 10.1103/PhysRevD.100.074011} {\bibfield  {journal} {\bibinfo
  {journal} {Phys. Rev. D}\ }\textbf {\bibinfo {volume} {100}},\ \bibinfo
  {pages} {074011} (\bibinfo {year} {2019})},\ \Eprint
  {http://arxiv.org/abs/1906.11644} {arXiv:1906.11644 [hep-ph]} \BibitemShut
  {NoStop}%
\bibitem [{\citenamefont {Fu}\ \emph {et~al.}(2020)\citenamefont {Fu},
  \citenamefont {Pawlowski},\ and\ \citenamefont {Rennecke}}]{Fu:2019hdw}%
  \BibitemOpen
  \bibfield  {author} {\bibinfo {author} {\bibfnamefont {W.-j.}\ \bibnamefont
  {Fu}}, \bibinfo {author} {\bibfnamefont {J.~M.}\ \bibnamefont {Pawlowski}}, \
  and\ \bibinfo {author} {\bibfnamefont {F.}~\bibnamefont {Rennecke}},\ }\href
  {\doibase 10.1103/PhysRevD.101.054032} {\bibfield  {journal} {\bibinfo
  {journal} {Phys. Rev. D}\ }\textbf {\bibinfo {volume} {101}},\ \bibinfo
  {pages} {054032} (\bibinfo {year} {2020})},\ \Eprint
  {http://arxiv.org/abs/1909.02991} {arXiv:1909.02991 [hep-ph]} \BibitemShut
  {NoStop}%
\bibitem [{\citenamefont {Gao}\ and\ \citenamefont
  {Pawlowski}(2020)}]{Gao:2020qsj}%
  \BibitemOpen
  \bibfield  {author} {\bibinfo {author} {\bibfnamefont {F.}~\bibnamefont
  {Gao}}\ and\ \bibinfo {author} {\bibfnamefont {J.~M.}\ \bibnamefont
  {Pawlowski}},\ }\href {\doibase 10.1103/PhysRevD.102.034027} {\bibfield
  {journal} {\bibinfo  {journal} {Phys. Rev. D}\ }\textbf {\bibinfo {volume}
  {102}},\ \bibinfo {pages} {034027} (\bibinfo {year} {2020})},\ \Eprint
  {http://arxiv.org/abs/2002.07500} {arXiv:2002.07500 [hep-ph]} \BibitemShut
  {NoStop}%
\bibitem [{\citenamefont {Gao}\ and\ \citenamefont
  {Pawlowski}(2021{\natexlab{a}})}]{Gao:2020fbl}%
  \BibitemOpen
  \bibfield  {author} {\bibinfo {author} {\bibfnamefont {F.}~\bibnamefont
  {Gao}}\ and\ \bibinfo {author} {\bibfnamefont {J.~M.}\ \bibnamefont
  {Pawlowski}},\ }\href {\doibase 10.1016/j.physletb.2021.136584} {\bibfield
  {journal} {\bibinfo  {journal} {Phys. Lett. B}\ }\textbf {\bibinfo {volume}
  {820}},\ \bibinfo {pages} {136584} (\bibinfo {year} {2021}{\natexlab{a}})},\
  \Eprint {http://arxiv.org/abs/2010.13705} {arXiv:2010.13705 [hep-ph]}
  \BibitemShut {NoStop}%
\bibitem [{\citenamefont {Gunkel}\ and\ \citenamefont
  {Fischer}(2021{\natexlab{a}})}]{Gunkel:2021oya}%
  \BibitemOpen
  \bibfield  {author} {\bibinfo {author} {\bibfnamefont {P.~J.}\ \bibnamefont
  {Gunkel}}\ and\ \bibinfo {author} {\bibfnamefont {C.~S.}\ \bibnamefont
  {Fischer}},\ }\href {\doibase 10.1103/PhysRevD.104.054022} {\bibfield
  {journal} {\bibinfo  {journal} {Phys. Rev. D}\ }\textbf {\bibinfo {volume}
  {104}},\ \bibinfo {pages} {054022} (\bibinfo {year} {2021}{\natexlab{a}})},\
  \Eprint {http://arxiv.org/abs/2106.08356} {arXiv:2106.08356 [hep-ph]}
  \BibitemShut {NoStop}%
\bibitem [{\citenamefont {Brown}\ \emph {et~al.}(1990)\citenamefont {Brown},
  \citenamefont {Butler}, \citenamefont {Chen}, \citenamefont {Christ},
  \citenamefont {Dong}, \citenamefont {Schaffer}, \citenamefont {Unger},\ and\
  \citenamefont {Vaccarino}}]{Brown:1990ev}%
  \BibitemOpen
  \bibfield  {author} {\bibinfo {author} {\bibfnamefont {F.~R.}\ \bibnamefont
  {Brown}}, \bibinfo {author} {\bibfnamefont {F.~P.}\ \bibnamefont {Butler}},
  \bibinfo {author} {\bibfnamefont {H.}~\bibnamefont {Chen}}, \bibinfo {author}
  {\bibfnamefont {N.~H.}\ \bibnamefont {Christ}}, \bibinfo {author}
  {\bibfnamefont {Z.-h.}\ \bibnamefont {Dong}}, \bibinfo {author}
  {\bibfnamefont {W.}~\bibnamefont {Schaffer}}, \bibinfo {author}
  {\bibfnamefont {L.~I.}\ \bibnamefont {Unger}}, \ and\ \bibinfo {author}
  {\bibfnamefont {A.}~\bibnamefont {Vaccarino}},\ }\href {\doibase
  10.1103/PhysRevLett.65.2491} {\bibfield  {journal} {\bibinfo  {journal}
  {Phys. Rev. Lett.}\ }\textbf {\bibinfo {volume} {65}},\ \bibinfo {pages}
  {2491} (\bibinfo {year} {1990})}\BibitemShut {NoStop}%
\bibitem [{\citenamefont {de~Forcrand}\ and\ \citenamefont
  {Philipsen}(2010)}]{deForcrand:2010he}%
  \BibitemOpen
  \bibfield  {author} {\bibinfo {author} {\bibfnamefont {P.}~\bibnamefont
  {de~Forcrand}}\ and\ \bibinfo {author} {\bibfnamefont {O.}~\bibnamefont
  {Philipsen}},\ }\href {\doibase 10.1103/PhysRevLett.105.152001} {\bibfield
  {journal} {\bibinfo  {journal} {Phys. Rev. Lett.}\ }\textbf {\bibinfo
  {volume} {105}},\ \bibinfo {pages} {152001} (\bibinfo {year} {2010})},\
  \Eprint {http://arxiv.org/abs/1004.3144} {arXiv:1004.3144 [hep-lat]}
  \BibitemShut {NoStop}%
\bibitem [{\citenamefont {Saito}\ \emph {et~al.}(2011)\citenamefont {Saito},
  \citenamefont {Ejiri}, \citenamefont {Aoki}, \citenamefont {Hatsuda},
  \citenamefont {Kanaya}, \citenamefont {Maezawa}, \citenamefont {Ohno},\ and\
  \citenamefont {Umeda}}]{Saito:2011fs}%
  \BibitemOpen
  \bibfield  {author} {\bibinfo {author} {\bibfnamefont {H.}~\bibnamefont
  {Saito}}, \bibinfo {author} {\bibfnamefont {S.}~\bibnamefont {Ejiri}},
  \bibinfo {author} {\bibfnamefont {S.}~\bibnamefont {Aoki}}, \bibinfo {author}
  {\bibfnamefont {T.}~\bibnamefont {Hatsuda}}, \bibinfo {author} {\bibfnamefont
  {K.}~\bibnamefont {Kanaya}}, \bibinfo {author} {\bibfnamefont
  {Y.}~\bibnamefont {Maezawa}}, \bibinfo {author} {\bibfnamefont
  {H.}~\bibnamefont {Ohno}}, \ and\ \bibinfo {author} {\bibfnamefont
  {T.}~\bibnamefont {Umeda}} (\bibinfo {collaboration} {WHOT-QCD}),\ }\href
  {\doibase 10.1103/PhysRevD.85.079902} {\bibfield  {journal} {\bibinfo
  {journal} {Phys. Rev. D}\ }\textbf {\bibinfo {volume} {84}},\ \bibinfo
  {pages} {054502} (\bibinfo {year} {2011})},\ \bibinfo {note} {[Erratum:
  Phys.Rev.D 85, 079902 (2012)]},\ \Eprint {http://arxiv.org/abs/1106.0974}
  {arXiv:1106.0974 [hep-lat]} \BibitemShut {NoStop}%
\bibitem [{\citenamefont {Fromm}\ \emph {et~al.}(2012)\citenamefont {Fromm},
  \citenamefont {Langelage}, \citenamefont {Lottini},\ and\ \citenamefont
  {Philipsen}}]{Fromm:2011qi}%
  \BibitemOpen
  \bibfield  {author} {\bibinfo {author} {\bibfnamefont {M.}~\bibnamefont
  {Fromm}}, \bibinfo {author} {\bibfnamefont {J.}~\bibnamefont {Langelage}},
  \bibinfo {author} {\bibfnamefont {S.}~\bibnamefont {Lottini}}, \ and\
  \bibinfo {author} {\bibfnamefont {O.}~\bibnamefont {Philipsen}},\ }\href
  {\doibase 10.1007/JHEP01(2012)042} {\bibfield  {journal} {\bibinfo  {journal}
  {JHEP}\ }\textbf {\bibinfo {volume} {01}},\ \bibinfo {pages} {042} (\bibinfo
  {year} {2012})},\ \Eprint {http://arxiv.org/abs/1111.4953} {arXiv:1111.4953
  [hep-lat]} \BibitemShut {NoStop}%
\bibitem [{\citenamefont {Ejiri}\ \emph {et~al.}(2020)\citenamefont {Ejiri},
  \citenamefont {Itagaki}, \citenamefont {Iwami}, \citenamefont {Kanaya},
  \citenamefont {Kitazawa}, \citenamefont {Kiyohara}, \citenamefont
  {Shirogane},\ and\ \citenamefont {Umeda}}]{Ejiri:2019csa}%
  \BibitemOpen
  \bibfield  {author} {\bibinfo {author} {\bibfnamefont {S.}~\bibnamefont
  {Ejiri}}, \bibinfo {author} {\bibfnamefont {S.}~\bibnamefont {Itagaki}},
  \bibinfo {author} {\bibfnamefont {R.}~\bibnamefont {Iwami}}, \bibinfo
  {author} {\bibfnamefont {K.}~\bibnamefont {Kanaya}}, \bibinfo {author}
  {\bibfnamefont {M.}~\bibnamefont {Kitazawa}}, \bibinfo {author}
  {\bibfnamefont {A.}~\bibnamefont {Kiyohara}}, \bibinfo {author}
  {\bibfnamefont {M.}~\bibnamefont {Shirogane}}, \ and\ \bibinfo {author}
  {\bibfnamefont {T.}~\bibnamefont {Umeda}} (\bibinfo {collaboration}
  {WHOT-QCD}),\ }\href {\doibase 10.1103/PhysRevD.101.054505} {\bibfield
  {journal} {\bibinfo  {journal} {Phys. Rev. D}\ }\textbf {\bibinfo {volume}
  {101}},\ \bibinfo {pages} {054505} (\bibinfo {year} {2020})},\ \Eprint
  {http://arxiv.org/abs/1912.10500} {arXiv:1912.10500 [hep-lat]} \BibitemShut
  {NoStop}%
\bibitem [{\citenamefont {Cuteri}\ \emph
  {et~al.}(2021{\natexlab{a}})\citenamefont {Cuteri}, \citenamefont
  {Philipsen}, \citenamefont {Sch\"on},\ and\ \citenamefont
  {Sciarra}}]{Cuteri:2020yke}%
  \BibitemOpen
  \bibfield  {author} {\bibinfo {author} {\bibfnamefont {F.}~\bibnamefont
  {Cuteri}}, \bibinfo {author} {\bibfnamefont {O.}~\bibnamefont {Philipsen}},
  \bibinfo {author} {\bibfnamefont {A.}~\bibnamefont {Sch\"on}}, \ and\
  \bibinfo {author} {\bibfnamefont {A.}~\bibnamefont {Sciarra}},\ }\href
  {\doibase 10.1103/PhysRevD.103.014513} {\bibfield  {journal} {\bibinfo
  {journal} {Phys. Rev. D}\ }\textbf {\bibinfo {volume} {103}},\ \bibinfo
  {pages} {014513} (\bibinfo {year} {2021}{\natexlab{a}})},\ \Eprint
  {http://arxiv.org/abs/2009.14033} {arXiv:2009.14033 [hep-lat]} \BibitemShut
  {NoStop}%
\bibitem [{\citenamefont {Kiyohara}\ \emph {et~al.}(2021)\citenamefont
  {Kiyohara}, \citenamefont {Kitazawa}, \citenamefont {Ejiri},\ and\
  \citenamefont {Kanaya}}]{Kiyohara:2021smr}%
  \BibitemOpen
  \bibfield  {author} {\bibinfo {author} {\bibfnamefont {A.}~\bibnamefont
  {Kiyohara}}, \bibinfo {author} {\bibfnamefont {M.}~\bibnamefont {Kitazawa}},
  \bibinfo {author} {\bibfnamefont {S.}~\bibnamefont {Ejiri}}, \ and\ \bibinfo
  {author} {\bibfnamefont {K.}~\bibnamefont {Kanaya}},\ }\href {\doibase
  10.1103/PhysRevD.104.114509} {\bibfield  {journal} {\bibinfo  {journal}
  {Phys. Rev. D}\ }\textbf {\bibinfo {volume} {104}},\ \bibinfo {pages}
  {114509} (\bibinfo {year} {2021})},\ \Eprint
  {http://arxiv.org/abs/2108.00118} {arXiv:2108.00118 [hep-lat]} \BibitemShut
  {NoStop}%
\bibitem [{\citenamefont {Kashiwa}\ \emph {et~al.}(2012)\citenamefont
  {Kashiwa}, \citenamefont {Pisarski},\ and\ \citenamefont
  {Skokov}}]{Kashiwa:2012wa}%
  \BibitemOpen
  \bibfield  {author} {\bibinfo {author} {\bibfnamefont {K.}~\bibnamefont
  {Kashiwa}}, \bibinfo {author} {\bibfnamefont {R.~D.}\ \bibnamefont
  {Pisarski}}, \ and\ \bibinfo {author} {\bibfnamefont {V.~V.}\ \bibnamefont
  {Skokov}},\ }\href {\doibase 10.1103/PhysRevD.85.114029} {\bibfield
  {journal} {\bibinfo  {journal} {Phys. Rev. D}\ }\textbf {\bibinfo {volume}
  {85}},\ \bibinfo {pages} {114029} (\bibinfo {year} {2012})},\ \Eprint
  {http://arxiv.org/abs/1205.0545} {arXiv:1205.0545 [hep-ph]} \BibitemShut
  {NoStop}%
\bibitem [{\citenamefont {Lo}\ \emph {et~al.}(2014)\citenamefont {Lo},
  \citenamefont {Friman},\ and\ \citenamefont {Redlich}}]{Lo:2014vba}%
  \BibitemOpen
  \bibfield  {author} {\bibinfo {author} {\bibfnamefont {P.~M.}\ \bibnamefont
  {Lo}}, \bibinfo {author} {\bibfnamefont {B.}~\bibnamefont {Friman}}, \ and\
  \bibinfo {author} {\bibfnamefont {K.}~\bibnamefont {Redlich}},\ }\href
  {\doibase 10.1103/PhysRevD.90.074035} {\bibfield  {journal} {\bibinfo
  {journal} {Phys. Rev. D}\ }\textbf {\bibinfo {volume} {90}},\ \bibinfo
  {pages} {074035} (\bibinfo {year} {2014})},\ \Eprint
  {http://arxiv.org/abs/1406.4050} {arXiv:1406.4050 [hep-ph]} \BibitemShut
  {NoStop}%
\bibitem [{\citenamefont {Fischer}\ \emph {et~al.}(2015)\citenamefont
  {Fischer}, \citenamefont {Luecker},\ and\ \citenamefont
  {Pawlowski}}]{Fischer:2014vxa}%
  \BibitemOpen
  \bibfield  {author} {\bibinfo {author} {\bibfnamefont {C.~S.}\ \bibnamefont
  {Fischer}}, \bibinfo {author} {\bibfnamefont {J.}~\bibnamefont {Luecker}}, \
  and\ \bibinfo {author} {\bibfnamefont {J.~M.}\ \bibnamefont {Pawlowski}},\
  }\href {\doibase 10.1103/PhysRevD.91.014024} {\bibfield  {journal} {\bibinfo
  {journal} {Phys. Rev. D}\ }\textbf {\bibinfo {volume} {91}},\ \bibinfo
  {pages} {014024} (\bibinfo {year} {2015})},\ \Eprint
  {http://arxiv.org/abs/1409.8462} {arXiv:1409.8462 [hep-ph]} \BibitemShut
  {NoStop}%
\bibitem [{\citenamefont {Reinosa}\ \emph {et~al.}(2015)\citenamefont
  {Reinosa}, \citenamefont {Serreau},\ and\ \citenamefont
  {Tissier}}]{Reinosa:2015oua}%
  \BibitemOpen
  \bibfield  {author} {\bibinfo {author} {\bibfnamefont {U.}~\bibnamefont
  {Reinosa}}, \bibinfo {author} {\bibfnamefont {J.}~\bibnamefont {Serreau}}, \
  and\ \bibinfo {author} {\bibfnamefont {M.}~\bibnamefont {Tissier}},\ }\href
  {\doibase 10.1103/PhysRevD.92.025021} {\bibfield  {journal} {\bibinfo
  {journal} {Phys. Rev. D}\ }\textbf {\bibinfo {volume} {92}},\ \bibinfo
  {pages} {025021} (\bibinfo {year} {2015})},\ \Eprint
  {http://arxiv.org/abs/1504.02916} {arXiv:1504.02916 [hep-th]} \BibitemShut
  {NoStop}%
\bibitem [{\citenamefont {Maelger}\ \emph {et~al.}(2018)\citenamefont
  {Maelger}, \citenamefont {Reinosa},\ and\ \citenamefont
  {Serreau}}]{Maelger:2017amh}%
  \BibitemOpen
  \bibfield  {author} {\bibinfo {author} {\bibfnamefont {J.}~\bibnamefont
  {Maelger}}, \bibinfo {author} {\bibfnamefont {U.}~\bibnamefont {Reinosa}}, \
  and\ \bibinfo {author} {\bibfnamefont {J.}~\bibnamefont {Serreau}},\ }\href
  {\doibase 10.1103/PhysRevD.97.074027} {\bibfield  {journal} {\bibinfo
  {journal} {Phys. Rev. D}\ }\textbf {\bibinfo {volume} {97}},\ \bibinfo
  {pages} {074027} (\bibinfo {year} {2018})},\ \Eprint
  {http://arxiv.org/abs/1710.01930} {arXiv:1710.01930 [hep-ph]} \BibitemShut
  {NoStop}%
\bibitem [{\citenamefont {Brandt}\ \emph {et~al.}(2016)\citenamefont {Brandt},
  \citenamefont {Francis}, \citenamefont {Meyer}, \citenamefont {Philipsen},
  \citenamefont {Robaina},\ and\ \citenamefont {Wittig}}]{Brandt:2016daq}%
  \BibitemOpen
  \bibfield  {author} {\bibinfo {author} {\bibfnamefont {B.~B.}\ \bibnamefont
  {Brandt}}, \bibinfo {author} {\bibfnamefont {A.}~\bibnamefont {Francis}},
  \bibinfo {author} {\bibfnamefont {H.~B.}\ \bibnamefont {Meyer}}, \bibinfo
  {author} {\bibfnamefont {O.}~\bibnamefont {Philipsen}}, \bibinfo {author}
  {\bibfnamefont {D.}~\bibnamefont {Robaina}}, \ and\ \bibinfo {author}
  {\bibfnamefont {H.}~\bibnamefont {Wittig}},\ }\href {\doibase
  10.1007/JHEP12(2016)158} {\bibfield  {journal} {\bibinfo  {journal} {JHEP}\
  }\textbf {\bibinfo {volume} {12}},\ \bibinfo {pages} {158} (\bibinfo {year}
  {2016})},\ \Eprint {http://arxiv.org/abs/1608.06882} {arXiv:1608.06882
  [hep-lat]} \BibitemShut {NoStop}%
\bibitem [{\citenamefont {Tomiya}\ \emph {et~al.}(2017)\citenamefont {Tomiya},
  \citenamefont {Cossu}, \citenamefont {Aoki}, \citenamefont {Fukaya},
  \citenamefont {Hashimoto}, \citenamefont {Kaneko},\ and\ \citenamefont
  {Noaki}}]{Tomiya:2016jwr}%
  \BibitemOpen
  \bibfield  {author} {\bibinfo {author} {\bibfnamefont {A.}~\bibnamefont
  {Tomiya}}, \bibinfo {author} {\bibfnamefont {G.}~\bibnamefont {Cossu}},
  \bibinfo {author} {\bibfnamefont {S.}~\bibnamefont {Aoki}}, \bibinfo {author}
  {\bibfnamefont {H.}~\bibnamefont {Fukaya}}, \bibinfo {author} {\bibfnamefont
  {S.}~\bibnamefont {Hashimoto}}, \bibinfo {author} {\bibfnamefont
  {T.}~\bibnamefont {Kaneko}}, \ and\ \bibinfo {author} {\bibfnamefont
  {J.}~\bibnamefont {Noaki}},\ }\href {\doibase 10.1103/PhysRevD.96.034509}
  {\bibfield  {journal} {\bibinfo  {journal} {Phys. Rev. D}\ }\textbf {\bibinfo
  {volume} {96}},\ \bibinfo {pages} {034509} (\bibinfo {year} {2017})},\
  \bibinfo {note} {[Addendum: Phys.Rev.D 96, 079902 (2017)]},\ \Eprint
  {http://arxiv.org/abs/1612.01908} {arXiv:1612.01908 [hep-lat]} \BibitemShut
  {NoStop}%
\bibitem [{\citenamefont {Aoki}\ \emph {et~al.}(2022)\citenamefont {Aoki},
  \citenamefont {Aoki}, \citenamefont {Fukaya}, \citenamefont {Hashimoto},
  \citenamefont {Rohrhofer},\ and\ \citenamefont {Suzuki}}]{Aoki:2021qws}%
  \BibitemOpen
  \bibfield  {author} {\bibinfo {author} {\bibfnamefont {S.}~\bibnamefont
  {Aoki}}, \bibinfo {author} {\bibfnamefont {Y.}~\bibnamefont {Aoki}}, \bibinfo
  {author} {\bibfnamefont {H.}~\bibnamefont {Fukaya}}, \bibinfo {author}
  {\bibfnamefont {S.}~\bibnamefont {Hashimoto}}, \bibinfo {author}
  {\bibfnamefont {C.}~\bibnamefont {Rohrhofer}}, \ and\ \bibinfo {author}
  {\bibfnamefont {K.}~\bibnamefont {Suzuki}} (\bibinfo {collaboration}
  {JLQCD}),\ }\href {\doibase 10.1093/ptep/ptac001} {\bibfield  {journal}
  {\bibinfo  {journal} {PTEP}\ }\textbf {\bibinfo {volume} {2022}},\ \bibinfo
  {pages} {023B05} (\bibinfo {year} {2022})},\ \Eprint
  {http://arxiv.org/abs/2103.05954} {arXiv:2103.05954 [hep-lat]} \BibitemShut
  {NoStop}%
\bibitem [{\citenamefont {Bazavov}\ \emph
  {et~al.}(2012{\natexlab{b}})\citenamefont {Bazavov} \emph
  {et~al.}}]{HotQCD:2012vvd}%
  \BibitemOpen
  \bibfield  {author} {\bibinfo {author} {\bibfnamefont {A.}~\bibnamefont
  {Bazavov}} \emph {et~al.} (\bibinfo {collaboration} {HotQCD}),\ }\href
  {\doibase 10.1103/PhysRevD.86.094503} {\bibfield  {journal} {\bibinfo
  {journal} {Phys. Rev. D}\ }\textbf {\bibinfo {volume} {86}},\ \bibinfo
  {pages} {094503} (\bibinfo {year} {2012}{\natexlab{b}})},\ \Eprint
  {http://arxiv.org/abs/1205.3535} {arXiv:1205.3535 [hep-lat]} \BibitemShut
  {NoStop}%
\bibitem [{\citenamefont {Buchoff}\ \emph {et~al.}(2014)\citenamefont {Buchoff}
  \emph {et~al.}}]{Buchoff:2013nra}%
  \BibitemOpen
  \bibfield  {author} {\bibinfo {author} {\bibfnamefont {M.~I.}\ \bibnamefont
  {Buchoff}} \emph {et~al.},\ }\href {\doibase 10.1103/PhysRevD.89.054514}
  {\bibfield  {journal} {\bibinfo  {journal} {Phys. Rev. D}\ }\textbf {\bibinfo
  {volume} {89}},\ \bibinfo {pages} {054514} (\bibinfo {year} {2014})},\
  \Eprint {http://arxiv.org/abs/1309.4149} {arXiv:1309.4149 [hep-lat]}
  \BibitemShut {NoStop}%
\bibitem [{\citenamefont {Bhattacharya}\ \emph {et~al.}(2014)\citenamefont
  {Bhattacharya} \emph {et~al.}}]{Bhattacharya:2014ara}%
  \BibitemOpen
  \bibfield  {author} {\bibinfo {author} {\bibfnamefont {T.}~\bibnamefont
  {Bhattacharya}} \emph {et~al.},\ }\href {\doibase
  10.1103/PhysRevLett.113.082001} {\bibfield  {journal} {\bibinfo  {journal}
  {Phys. Rev. Lett.}\ }\textbf {\bibinfo {volume} {113}},\ \bibinfo {pages}
  {082001} (\bibinfo {year} {2014})},\ \Eprint {http://arxiv.org/abs/1402.5175}
  {arXiv:1402.5175 [hep-lat]} \BibitemShut {NoStop}%
\bibitem [{\citenamefont {Dick}\ \emph {et~al.}(2015)\citenamefont {Dick},
  \citenamefont {Karsch}, \citenamefont {Laermann}, \citenamefont {Mukherjee},\
  and\ \citenamefont {Sharma}}]{Dick:2015twa}%
  \BibitemOpen
  \bibfield  {author} {\bibinfo {author} {\bibfnamefont {V.}~\bibnamefont
  {Dick}}, \bibinfo {author} {\bibfnamefont {F.}~\bibnamefont {Karsch}},
  \bibinfo {author} {\bibfnamefont {E.}~\bibnamefont {Laermann}}, \bibinfo
  {author} {\bibfnamefont {S.}~\bibnamefont {Mukherjee}}, \ and\ \bibinfo
  {author} {\bibfnamefont {S.}~\bibnamefont {Sharma}},\ }\href {\doibase
  10.1103/PhysRevD.91.094504} {\bibfield  {journal} {\bibinfo  {journal} {Phys.
  Rev.}\ }\textbf {\bibinfo {volume} {D91}},\ \bibinfo {pages} {094504}
  (\bibinfo {year} {2015})},\ \Eprint {http://arxiv.org/abs/1502.06190}
  {arXiv:1502.06190 [hep-lat]} \BibitemShut {NoStop}%
\bibitem [{\citenamefont {Ding}\ \emph {et~al.}(2021)\citenamefont {Ding},
  \citenamefont {Li}, \citenamefont {Mukherjee}, \citenamefont {Tomiya},
  \citenamefont {Wang},\ and\ \citenamefont {Zhang}}]{Ding:2020xlj}%
  \BibitemOpen
  \bibfield  {author} {\bibinfo {author} {\bibfnamefont {H.~T.}\ \bibnamefont
  {Ding}}, \bibinfo {author} {\bibfnamefont {S.~T.}\ \bibnamefont {Li}},
  \bibinfo {author} {\bibfnamefont {S.}~\bibnamefont {Mukherjee}}, \bibinfo
  {author} {\bibfnamefont {A.}~\bibnamefont {Tomiya}}, \bibinfo {author}
  {\bibfnamefont {X.~D.}\ \bibnamefont {Wang}}, \ and\ \bibinfo {author}
  {\bibfnamefont {Y.}~\bibnamefont {Zhang}},\ }\href {\doibase
  10.1103/PhysRevLett.126.082001} {\bibfield  {journal} {\bibinfo  {journal}
  {Phys. Rev. Lett.}\ }\textbf {\bibinfo {volume} {126}},\ \bibinfo {pages}
  {082001} (\bibinfo {year} {2021})},\ \Eprint
  {http://arxiv.org/abs/2010.14836} {arXiv:2010.14836 [hep-lat]} \BibitemShut
  {NoStop}%
\bibitem [{\citenamefont {Kaczmarek}\ \emph {et~al.}(2021)\citenamefont
  {Kaczmarek}, \citenamefont {Mazur},\ and\ \citenamefont
  {Sharma}}]{Kaczmarek:2021ser}%
  \BibitemOpen
  \bibfield  {author} {\bibinfo {author} {\bibfnamefont {O.}~\bibnamefont
  {Kaczmarek}}, \bibinfo {author} {\bibfnamefont {L.}~\bibnamefont {Mazur}}, \
  and\ \bibinfo {author} {\bibfnamefont {S.}~\bibnamefont {Sharma}},\ }\href
  {\doibase 10.1103/PhysRevD.104.094518} {\bibfield  {journal} {\bibinfo
  {journal} {Phys. Rev. D}\ }\textbf {\bibinfo {volume} {104}},\ \bibinfo
  {pages} {094518} (\bibinfo {year} {2021})},\ \Eprint
  {http://arxiv.org/abs/2102.06136} {arXiv:2102.06136 [hep-lat]} \BibitemShut
  {NoStop}%
\bibitem [{\citenamefont {Pisarski}\ and\ \citenamefont
  {Wilczek}(1984)}]{Pisarski:1983ms}%
  \BibitemOpen
  \bibfield  {author} {\bibinfo {author} {\bibfnamefont {R.~D.}\ \bibnamefont
  {Pisarski}}\ and\ \bibinfo {author} {\bibfnamefont {F.}~\bibnamefont
  {Wilczek}},\ }\href {\doibase 10.1103/PhysRevD.29.338} {\bibfield  {journal}
  {\bibinfo  {journal} {Phys. Rev. D}\ }\textbf {\bibinfo {volume} {29}},\
  \bibinfo {pages} {338} (\bibinfo {year} {1984})}\BibitemShut {NoStop}%
\bibitem [{\citenamefont {Butti}\ \emph {et~al.}(2003)\citenamefont {Butti},
  \citenamefont {Pelissetto},\ and\ \citenamefont {Vicari}}]{Butti:2003nu}%
  \BibitemOpen
  \bibfield  {author} {\bibinfo {author} {\bibfnamefont {A.}~\bibnamefont
  {Butti}}, \bibinfo {author} {\bibfnamefont {A.}~\bibnamefont {Pelissetto}}, \
  and\ \bibinfo {author} {\bibfnamefont {E.}~\bibnamefont {Vicari}},\ }\href
  {\doibase 10.1088/1126-6708/2003/08/029} {\bibfield  {journal} {\bibinfo
  {journal} {JHEP}\ }\textbf {\bibinfo {volume} {08}},\ \bibinfo {pages} {029}
  (\bibinfo {year} {2003})},\ \Eprint {http://arxiv.org/abs/hep-ph/0307036}
  {arXiv:hep-ph/0307036 [hep-ph]} \BibitemShut {NoStop}%
\bibitem [{\citenamefont {de~Forcrand}\ and\ \citenamefont
  {D'Elia}(2017)}]{deForcrand:2017cgb}%
  \BibitemOpen
  \bibfield  {author} {\bibinfo {author} {\bibfnamefont {P.}~\bibnamefont
  {de~Forcrand}}\ and\ \bibinfo {author} {\bibfnamefont {M.}~\bibnamefont
  {D'Elia}},\ }\bibfield  {booktitle} {\emph {\bibinfo {booktitle}
  {{Proceedings, 34th International Symposium on Lattice Field Theory (Lattice
  2016): Southampton, UK, July 24-30, 2016}}},\ }\href@noop {} {\bibfield
  {journal} {\bibinfo  {journal} {PoS}\ }\textbf {\bibinfo {volume}
  {LATTICE2016}},\ \bibinfo {pages} {081} (\bibinfo {year} {2017})},\ \Eprint
  {http://arxiv.org/abs/1702.00330} {arXiv:1702.00330 [hep-lat]} \BibitemShut
  {NoStop}%
\bibitem [{\citenamefont {Iwasaki}\ \emph {et~al.}(1997)\citenamefont
  {Iwasaki}, \citenamefont {Kanaya}, \citenamefont {Kaya},\ and\ \citenamefont
  {Yoshie}}]{Iwasaki:1996ya}%
  \BibitemOpen
  \bibfield  {author} {\bibinfo {author} {\bibfnamefont {Y.}~\bibnamefont
  {Iwasaki}}, \bibinfo {author} {\bibfnamefont {K.}~\bibnamefont {Kanaya}},
  \bibinfo {author} {\bibfnamefont {S.}~\bibnamefont {Kaya}}, \ and\ \bibinfo
  {author} {\bibfnamefont {T.}~\bibnamefont {Yoshie}},\ }\href {\doibase
  10.1103/PhysRevLett.78.179} {\bibfield  {journal} {\bibinfo  {journal} {Phys.
  Rev. Lett.}\ }\textbf {\bibinfo {volume} {78}},\ \bibinfo {pages} {179}
  (\bibinfo {year} {1997})},\ \Eprint {http://arxiv.org/abs/hep-lat/9609022}
  {arXiv:hep-lat/9609022 [hep-lat]} \BibitemShut {NoStop}%
\bibitem [{\citenamefont {D'Elia}\ \emph
  {et~al.}(2005{\natexlab{a}})\citenamefont {D'Elia}, \citenamefont
  {Di~Giacomo},\ and\ \citenamefont {Pica}}]{DElia:2004uwa}%
  \BibitemOpen
  \bibfield  {author} {\bibinfo {author} {\bibfnamefont {M.}~\bibnamefont
  {D'Elia}}, \bibinfo {author} {\bibfnamefont {A.}~\bibnamefont {Di~Giacomo}},
  \ and\ \bibinfo {author} {\bibfnamefont {C.}~\bibnamefont {Pica}},\
  }\bibfield  {booktitle} {\emph {\bibinfo {booktitle} {{Non-perturbative
  quantum chromodynamics. Proceedings, 8th Workshop, Paris, France, June 7-11,
  2004}}},\ }\href {\doibase 10.1142/S0217751X05028235} {\bibfield  {journal}
  {\bibinfo  {journal} {Int. J. Mod. Phys.}\ }\textbf {\bibinfo {volume}
  {A20}},\ \bibinfo {pages} {4579} (\bibinfo {year} {2005}{\natexlab{a}})},\
  \Eprint {http://arxiv.org/abs/hep-lat/0408011} {arXiv:hep-lat/0408011
  [hep-lat]} \BibitemShut {NoStop}%
\bibitem [{\citenamefont {D'Elia}\ \emph
  {et~al.}(2005{\natexlab{b}})\citenamefont {D'Elia}, \citenamefont
  {Di~Giacomo},\ and\ \citenamefont {Pica}}]{DElia:2005nmv}%
  \BibitemOpen
  \bibfield  {author} {\bibinfo {author} {\bibfnamefont {M.}~\bibnamefont
  {D'Elia}}, \bibinfo {author} {\bibfnamefont {A.}~\bibnamefont {Di~Giacomo}},
  \ and\ \bibinfo {author} {\bibfnamefont {C.}~\bibnamefont {Pica}},\ }\href
  {\doibase 10.1103/PhysRevD.72.114510} {\bibfield  {journal} {\bibinfo
  {journal} {Phys. Rev.}\ }\textbf {\bibinfo {volume} {D72}},\ \bibinfo {pages}
  {114510} (\bibinfo {year} {2005}{\natexlab{b}})},\ \Eprint
  {http://arxiv.org/abs/hep-lat/0503030} {arXiv:hep-lat/0503030 [hep-lat]}
  \BibitemShut {NoStop}%
\bibitem [{\citenamefont {Kogut}\ and\ \citenamefont
  {Sinclair}(2006)}]{Kogut:2006gt}%
  \BibitemOpen
  \bibfield  {author} {\bibinfo {author} {\bibfnamefont {J.~B.}\ \bibnamefont
  {Kogut}}\ and\ \bibinfo {author} {\bibfnamefont {D.~K.}\ \bibnamefont
  {Sinclair}},\ }\href {\doibase 10.1103/PhysRevD.73.074512} {\bibfield
  {journal} {\bibinfo  {journal} {Phys. Rev.}\ }\textbf {\bibinfo {volume}
  {D73}},\ \bibinfo {pages} {074512} (\bibinfo {year} {2006})},\ \Eprint
  {http://arxiv.org/abs/hep-lat/0603021} {arXiv:hep-lat/0603021 [hep-lat]}
  \BibitemShut {NoStop}%
\bibitem [{\citenamefont {Bonati}\ \emph {et~al.}(2014)\citenamefont {Bonati},
  \citenamefont {de~Forcrand}, \citenamefont {D'Elia}, \citenamefont
  {Philipsen},\ and\ \citenamefont {Sanfilippo}}]{Bonati:2014kpa}%
  \BibitemOpen
  \bibfield  {author} {\bibinfo {author} {\bibfnamefont {C.}~\bibnamefont
  {Bonati}}, \bibinfo {author} {\bibfnamefont {P.}~\bibnamefont {de~Forcrand}},
  \bibinfo {author} {\bibfnamefont {M.}~\bibnamefont {D'Elia}}, \bibinfo
  {author} {\bibfnamefont {O.}~\bibnamefont {Philipsen}}, \ and\ \bibinfo
  {author} {\bibfnamefont {F.}~\bibnamefont {Sanfilippo}},\ }\href {\doibase
  10.1103/PhysRevD.90.074030} {\bibfield  {journal} {\bibinfo  {journal} {Phys.
  Rev.}\ }\textbf {\bibinfo {volume} {D90}},\ \bibinfo {pages} {074030}
  (\bibinfo {year} {2014})},\ \Eprint {http://arxiv.org/abs/1408.5086}
  {arXiv:1408.5086 [hep-lat]} \BibitemShut {NoStop}%
\bibitem [{\citenamefont {Philipsen}\ and\ \citenamefont
  {Pinke}(2016)}]{Philipsen:2016hkv}%
  \BibitemOpen
  \bibfield  {author} {\bibinfo {author} {\bibfnamefont {O.}~\bibnamefont
  {Philipsen}}\ and\ \bibinfo {author} {\bibfnamefont {C.}~\bibnamefont
  {Pinke}},\ }\href {\doibase 10.1103/PhysRevD.93.114507} {\bibfield  {journal}
  {\bibinfo  {journal} {Phys. Rev.}\ }\textbf {\bibinfo {volume} {D93}},\
  \bibinfo {pages} {114507} (\bibinfo {year} {2016})},\ \Eprint
  {http://arxiv.org/abs/1602.06129} {arXiv:1602.06129 [hep-lat]} \BibitemShut
  {NoStop}%
\bibitem [{\citenamefont {Cuteri}\ \emph {et~al.}(2018)\citenamefont {Cuteri},
  \citenamefont {Philipsen},\ and\ \citenamefont {Sciarra}}]{Cuteri:2017gci}%
  \BibitemOpen
  \bibfield  {author} {\bibinfo {author} {\bibfnamefont {F.}~\bibnamefont
  {Cuteri}}, \bibinfo {author} {\bibfnamefont {O.}~\bibnamefont {Philipsen}}, \
  and\ \bibinfo {author} {\bibfnamefont {A.}~\bibnamefont {Sciarra}},\ }\href
  {\doibase 10.1103/PhysRevD.97.114511} {\bibfield  {journal} {\bibinfo
  {journal} {Phys. Rev.}\ }\textbf {\bibinfo {volume} {D97}},\ \bibinfo {pages}
  {114511} (\bibinfo {year} {2018})},\ \Eprint
  {http://arxiv.org/abs/1711.05658} {arXiv:1711.05658 [hep-lat]} \BibitemShut
  {NoStop}%
\bibitem [{\citenamefont {Ding}\ \emph {et~al.}(2018)\citenamefont {Ding},
  \citenamefont {Hegde}, \citenamefont {Karsch}, \citenamefont {Lahiri},
  \citenamefont {Li}, \citenamefont {Mukherjee},\ and\ \citenamefont
  {Petreczky}}]{Ding:2018auz}%
  \BibitemOpen
  \bibfield  {author} {\bibinfo {author} {\bibfnamefont {H.~T.}\ \bibnamefont
  {Ding}}, \bibinfo {author} {\bibfnamefont {P.}~\bibnamefont {Hegde}},
  \bibinfo {author} {\bibfnamefont {F.}~\bibnamefont {Karsch}}, \bibinfo
  {author} {\bibfnamefont {A.}~\bibnamefont {Lahiri}}, \bibinfo {author}
  {\bibfnamefont {S.~T.}\ \bibnamefont {Li}}, \bibinfo {author} {\bibfnamefont
  {S.}~\bibnamefont {Mukherjee}}, \ and\ \bibinfo {author} {\bibfnamefont
  {P.}~\bibnamefont {Petreczky}},\ }in\ \href@noop {} {\emph {\bibinfo
  {booktitle} {{27th International Conference on Ultrarelativistic
  Nucleus-Nucleus Collisions (Quark Matter 2018) Venice, Italy, May 14-19,
  2018}}}}\ (\bibinfo {year} {2018})\ \Eprint {http://arxiv.org/abs/1807.05727}
  {arXiv:1807.05727 [hep-lat]} \BibitemShut {NoStop}%
\bibitem [{\citenamefont {Lenaghan}(2001)}]{Lenaghan:2000kr}%
  \BibitemOpen
  \bibfield  {author} {\bibinfo {author} {\bibfnamefont {J.~T.}\ \bibnamefont
  {Lenaghan}},\ }\href {\doibase 10.1103/PhysRevD.63.037901} {\bibfield
  {journal} {\bibinfo  {journal} {Phys. Rev.}\ }\textbf {\bibinfo {volume}
  {D63}},\ \bibinfo {pages} {037901} (\bibinfo {year} {2001})},\ \Eprint
  {http://arxiv.org/abs/hep-ph/0005330} {arXiv:hep-ph/0005330 [hep-ph]}
  \BibitemShut {NoStop}%
\bibitem [{\citenamefont {Kovacs}\ and\ \citenamefont
  {Szep}(2007)}]{Kovacs:2006ym}%
  \BibitemOpen
  \bibfield  {author} {\bibinfo {author} {\bibfnamefont {P.}~\bibnamefont
  {Kovacs}}\ and\ \bibinfo {author} {\bibfnamefont {Z.}~\bibnamefont {Szep}},\
  }\href {\doibase 10.1103/PhysRevD.75.025015} {\bibfield  {journal} {\bibinfo
  {journal} {Phys. Rev.}\ }\textbf {\bibinfo {volume} {D75}},\ \bibinfo {pages}
  {025015} (\bibinfo {year} {2007})},\ \Eprint
  {http://arxiv.org/abs/hep-ph/0611208} {arXiv:hep-ph/0611208 [hep-ph]}
  \BibitemShut {NoStop}%
\bibitem [{\citenamefont {Fukushima}(2008)}]{Fukushima:2008wg}%
  \BibitemOpen
  \bibfield  {author} {\bibinfo {author} {\bibfnamefont {K.}~\bibnamefont
  {Fukushima}},\ }\href {\doibase 10.1103/PhysRevD.77.114028,
  10.1103/PhysRevD.78.039902} {\bibfield  {journal} {\bibinfo  {journal} {Phys.
  Rev.}\ }\textbf {\bibinfo {volume} {D77}},\ \bibinfo {pages} {114028}
  (\bibinfo {year} {2008})},\ \bibinfo {note} {[Erratum: Phys.
  Rev.D78,039902(2008)]},\ \Eprint {http://arxiv.org/abs/0803.3318}
  {arXiv:0803.3318 [hep-ph]} \BibitemShut {NoStop}%
\bibitem [{\citenamefont {Schaefer}\ and\ \citenamefont
  {Wagner}(2009)}]{Schaefer:2008hk}%
  \BibitemOpen
  \bibfield  {author} {\bibinfo {author} {\bibfnamefont {B.-J.}\ \bibnamefont
  {Schaefer}}\ and\ \bibinfo {author} {\bibfnamefont {M.}~\bibnamefont
  {Wagner}},\ }\href {\doibase 10.1103/PhysRevD.79.014018} {\bibfield
  {journal} {\bibinfo  {journal} {Phys. Rev.}\ }\textbf {\bibinfo {volume}
  {D79}},\ \bibinfo {pages} {014018} (\bibinfo {year} {2009})},\ \Eprint
  {http://arxiv.org/abs/0808.1491} {arXiv:0808.1491 [hep-ph]} \BibitemShut
  {NoStop}%
\bibitem [{\citenamefont {Mitter}\ and\ \citenamefont
  {Schaefer}(2014)}]{Mitter:2013fxa}%
  \BibitemOpen
  \bibfield  {author} {\bibinfo {author} {\bibfnamefont {M.}~\bibnamefont
  {Mitter}}\ and\ \bibinfo {author} {\bibfnamefont {B.-J.}\ \bibnamefont
  {Schaefer}},\ }\href {\doibase 10.1103/PhysRevD.89.054027} {\bibfield
  {journal} {\bibinfo  {journal} {Phys. Rev.}\ }\textbf {\bibinfo {volume}
  {D89}},\ \bibinfo {pages} {054027} (\bibinfo {year} {2014})},\ \Eprint
  {http://arxiv.org/abs/1308.3176} {arXiv:1308.3176 [hep-ph]} \BibitemShut
  {NoStop}%
\bibitem [{\citenamefont {Grahl}\ and\ \citenamefont
  {Rischke}(2013)}]{Grahl:2013pba}%
  \BibitemOpen
  \bibfield  {author} {\bibinfo {author} {\bibfnamefont {M.}~\bibnamefont
  {Grahl}}\ and\ \bibinfo {author} {\bibfnamefont {D.~H.}\ \bibnamefont
  {Rischke}},\ }\href {\doibase 10.1103/PhysRevD.88.056014} {\bibfield
  {journal} {\bibinfo  {journal} {Phys. Rev. D}\ }\textbf {\bibinfo {volume}
  {88}},\ \bibinfo {pages} {056014} (\bibinfo {year} {2013})},\ \Eprint
  {http://arxiv.org/abs/1307.2184} {arXiv:1307.2184 [hep-th]} \BibitemShut
  {NoStop}%
\bibitem [{\citenamefont {Eser}\ \emph {et~al.}(2015)\citenamefont {Eser},
  \citenamefont {Grahl},\ and\ \citenamefont {Rischke}}]{Eser:2015pka}%
  \BibitemOpen
  \bibfield  {author} {\bibinfo {author} {\bibfnamefont {J.}~\bibnamefont
  {Eser}}, \bibinfo {author} {\bibfnamefont {M.}~\bibnamefont {Grahl}}, \ and\
  \bibinfo {author} {\bibfnamefont {D.~H.}\ \bibnamefont {Rischke}},\ }\href
  {\doibase 10.1103/PhysRevD.92.096008} {\bibfield  {journal} {\bibinfo
  {journal} {Phys. Rev. D}\ }\textbf {\bibinfo {volume} {92}},\ \bibinfo
  {pages} {096008} (\bibinfo {year} {2015})},\ \Eprint
  {http://arxiv.org/abs/1508.06928} {arXiv:1508.06928 [hep-ph]} \BibitemShut
  {NoStop}%
\bibitem [{\citenamefont {Resch}\ \emph {et~al.}(2017)\citenamefont {Resch},
  \citenamefont {Rennecke},\ and\ \citenamefont {Schaefer}}]{Resch:2017vjs}%
  \BibitemOpen
  \bibfield  {author} {\bibinfo {author} {\bibfnamefont {S.}~\bibnamefont
  {Resch}}, \bibinfo {author} {\bibfnamefont {F.}~\bibnamefont {Rennecke}}, \
  and\ \bibinfo {author} {\bibfnamefont {B.-J.}\ \bibnamefont {Schaefer}},\
  }\href@noop {} {\  (\bibinfo {year} {2017})},\ \Eprint
  {http://arxiv.org/abs/1712.07961} {arXiv:1712.07961 [hep-ph]} \BibitemShut
  {NoStop}%
\bibitem [{\citenamefont {Braun}\ \emph {et~al.}(2011)\citenamefont {Braun},
  \citenamefont {Haas}, \citenamefont {Marhauser},\ and\ \citenamefont
  {Pawlowski}}]{Braun:2009gm}%
  \BibitemOpen
  \bibfield  {author} {\bibinfo {author} {\bibfnamefont {J.}~\bibnamefont
  {Braun}}, \bibinfo {author} {\bibfnamefont {L.~M.}\ \bibnamefont {Haas}},
  \bibinfo {author} {\bibfnamefont {F.}~\bibnamefont {Marhauser}}, \ and\
  \bibinfo {author} {\bibfnamefont {J.~M.}\ \bibnamefont {Pawlowski}},\ }\href
  {\doibase 10.1103/PhysRevLett.106.022002} {\bibfield  {journal} {\bibinfo
  {journal} {Phys. Rev. Lett.}\ }\textbf {\bibinfo {volume} {106}},\ \bibinfo
  {pages} {022002} (\bibinfo {year} {2011})},\ \Eprint
  {http://arxiv.org/abs/0908.0008} {arXiv:0908.0008 [hep-ph]} \BibitemShut
  {NoStop}%
\bibitem [{\citenamefont {Braun}\ \emph {et~al.}(2020)\citenamefont {Braun},
  \citenamefont {Fu}, \citenamefont {Pawlowski}, \citenamefont {Rennecke},
  \citenamefont {Rosenbl\"uh},\ and\ \citenamefont {Yin}}]{Braun:2020ada}%
  \BibitemOpen
  \bibfield  {author} {\bibinfo {author} {\bibfnamefont {J.}~\bibnamefont
  {Braun}}, \bibinfo {author} {\bibfnamefont {W.-j.}\ \bibnamefont {Fu}},
  \bibinfo {author} {\bibfnamefont {J.~M.}\ \bibnamefont {Pawlowski}}, \bibinfo
  {author} {\bibfnamefont {F.}~\bibnamefont {Rennecke}}, \bibinfo {author}
  {\bibfnamefont {D.}~\bibnamefont {Rosenbl\"uh}}, \ and\ \bibinfo {author}
  {\bibfnamefont {S.}~\bibnamefont {Yin}},\ }\href {\doibase
  10.1103/PhysRevD.102.056010} {\bibfield  {journal} {\bibinfo  {journal}
  {Phys. Rev. D}\ }\textbf {\bibinfo {volume} {102}},\ \bibinfo {pages}
  {056010} (\bibinfo {year} {2020})},\ \Eprint
  {http://arxiv.org/abs/2003.13112} {arXiv:2003.13112 [hep-ph]} \BibitemShut
  {NoStop}%
\bibitem [{\citenamefont {Karsch}\ \emph {et~al.}(2001)\citenamefont {Karsch},
  \citenamefont {Laermann},\ and\ \citenamefont {Schmidt}}]{Karsch:2001nf}%
  \BibitemOpen
  \bibfield  {author} {\bibinfo {author} {\bibfnamefont {F.}~\bibnamefont
  {Karsch}}, \bibinfo {author} {\bibfnamefont {E.}~\bibnamefont {Laermann}}, \
  and\ \bibinfo {author} {\bibfnamefont {C.}~\bibnamefont {Schmidt}},\ }\href
  {\doibase 10.1016/S0370-2693(01)01114-5} {\bibfield  {journal} {\bibinfo
  {journal} {Phys. Lett.}\ }\textbf {\bibinfo {volume} {B520}},\ \bibinfo
  {pages} {41} (\bibinfo {year} {2001})},\ \Eprint
  {http://arxiv.org/abs/hep-lat/0107020} {arXiv:hep-lat/0107020 [hep-lat]}
  \BibitemShut {NoStop}%
\bibitem [{\citenamefont {Karsch}\ \emph {et~al.}(2004)\citenamefont {Karsch},
  \citenamefont {Allton}, \citenamefont {Ejiri}, \citenamefont {Hands},
  \citenamefont {Kaczmarek}, \citenamefont {Laermann},\ and\ \citenamefont
  {Schmidt}}]{Karsch:2003va}%
  \BibitemOpen
  \bibfield  {author} {\bibinfo {author} {\bibfnamefont {F.}~\bibnamefont
  {Karsch}}, \bibinfo {author} {\bibfnamefont {C.~R.}\ \bibnamefont {Allton}},
  \bibinfo {author} {\bibfnamefont {S.}~\bibnamefont {Ejiri}}, \bibinfo
  {author} {\bibfnamefont {S.~J.}\ \bibnamefont {Hands}}, \bibinfo {author}
  {\bibfnamefont {O.}~\bibnamefont {Kaczmarek}}, \bibinfo {author}
  {\bibfnamefont {E.}~\bibnamefont {Laermann}}, \ and\ \bibinfo {author}
  {\bibfnamefont {C.}~\bibnamefont {Schmidt}},\ }\bibfield  {booktitle} {\emph
  {\bibinfo {booktitle} {{Lattice field theory. Proceedings, 21st International
  Symposium, Lattice 2003, Tsukuba, Japan, July 15-19, 2003}}},\ }\href
  {\doibase 10.1016/S0920-5632(03)02659-8} {\bibfield  {journal} {\bibinfo
  {journal} {Nucl. Phys. Proc. Suppl.}\ }\textbf {\bibinfo {volume} {129}},\
  \bibinfo {pages} {614} (\bibinfo {year} {2004})},\ \bibinfo {note}
  {[,614(2003)]},\ \Eprint {http://arxiv.org/abs/hep-lat/0309116}
  {arXiv:hep-lat/0309116 [hep-lat]} \BibitemShut {NoStop}%
\bibitem [{\citenamefont {de~Forcrand}\ \emph {et~al.}(2007)\citenamefont
  {de~Forcrand}, \citenamefont {Kim},\ and\ \citenamefont
  {Philipsen}}]{deForcrand:2007rq}%
  \BibitemOpen
  \bibfield  {author} {\bibinfo {author} {\bibfnamefont {P.}~\bibnamefont
  {de~Forcrand}}, \bibinfo {author} {\bibfnamefont {S.}~\bibnamefont {Kim}}, \
  and\ \bibinfo {author} {\bibfnamefont {O.}~\bibnamefont {Philipsen}},\
  }\bibfield  {booktitle} {\emph {\bibinfo {booktitle} {{Proceedings, 25th
  International Symposium on Lattice field theory (Lattice 2007): Regensburg,
  Germany, July 30-August 4, 2007}}},\ }\href@noop {} {\bibfield  {journal}
  {\bibinfo  {journal} {PoS}\ }\textbf {\bibinfo {volume} {LATTICE2007}},\
  \bibinfo {pages} {178} (\bibinfo {year} {2007})},\ \Eprint
  {http://arxiv.org/abs/0711.0262} {arXiv:0711.0262 [hep-lat]} \BibitemShut
  {NoStop}%
\bibitem [{\citenamefont {Ding}\ \emph {et~al.}(2011)\citenamefont {Ding},
  \citenamefont {Bazavov}, \citenamefont {Hegde}, \citenamefont {Karsch},
  \citenamefont {Mukherjee},\ and\ \citenamefont {Petreczky}}]{Ding:2011du}%
  \BibitemOpen
  \bibfield  {author} {\bibinfo {author} {\bibfnamefont {H.~T.}\ \bibnamefont
  {Ding}}, \bibinfo {author} {\bibfnamefont {A.}~\bibnamefont {Bazavov}},
  \bibinfo {author} {\bibfnamefont {P.}~\bibnamefont {Hegde}}, \bibinfo
  {author} {\bibfnamefont {F.}~\bibnamefont {Karsch}}, \bibinfo {author}
  {\bibfnamefont {S.}~\bibnamefont {Mukherjee}}, \ and\ \bibinfo {author}
  {\bibfnamefont {P.}~\bibnamefont {Petreczky}},\ }\bibfield  {booktitle}
  {\emph {\bibinfo {booktitle} {{Proceedings, 29th International Symposium on
  Lattice field theory (Lattice 2011): Squaw Valley, Lake Tahoe, USA, July
  10-16, 2011}}},\ }\href@noop {} {\bibfield  {journal} {\bibinfo  {journal}
  {PoS}\ }\textbf {\bibinfo {volume} {LATTICE2011}},\ \bibinfo {pages} {191}
  (\bibinfo {year} {2011})},\ \Eprint {http://arxiv.org/abs/1111.0185}
  {arXiv:1111.0185 [hep-lat]} \BibitemShut {NoStop}%
\bibitem [{\citenamefont {Jin}\ \emph {et~al.}(2015)\citenamefont {Jin},
  \citenamefont {Kuramashi}, \citenamefont {Nakamura}, \citenamefont {Takeda},\
  and\ \citenamefont {Ukawa}}]{Jin:2014hea}%
  \BibitemOpen
  \bibfield  {author} {\bibinfo {author} {\bibfnamefont {X.-Y.}\ \bibnamefont
  {Jin}}, \bibinfo {author} {\bibfnamefont {Y.}~\bibnamefont {Kuramashi}},
  \bibinfo {author} {\bibfnamefont {Y.}~\bibnamefont {Nakamura}}, \bibinfo
  {author} {\bibfnamefont {S.}~\bibnamefont {Takeda}}, \ and\ \bibinfo {author}
  {\bibfnamefont {A.}~\bibnamefont {Ukawa}},\ }\href {\doibase
  10.1103/PhysRevD.91.014508} {\bibfield  {journal} {\bibinfo  {journal} {Phys.
  Rev.}\ }\textbf {\bibinfo {volume} {D91}},\ \bibinfo {pages} {014508}
  (\bibinfo {year} {2015})},\ \Eprint {http://arxiv.org/abs/1411.7461}
  {arXiv:1411.7461 [hep-lat]} \BibitemShut {NoStop}%
\bibitem [{\citenamefont {Takeda}\ \emph {et~al.}(2017)\citenamefont {Takeda},
  \citenamefont {Jin}, \citenamefont {Kuramashi}, \citenamefont {Nakamura},\
  and\ \citenamefont {Ukawa}}]{Takeda:2016vfj}%
  \BibitemOpen
  \bibfield  {author} {\bibinfo {author} {\bibfnamefont {S.}~\bibnamefont
  {Takeda}}, \bibinfo {author} {\bibfnamefont {X.-Y.}\ \bibnamefont {Jin}},
  \bibinfo {author} {\bibfnamefont {Y.}~\bibnamefont {Kuramashi}}, \bibinfo
  {author} {\bibfnamefont {Y.}~\bibnamefont {Nakamura}}, \ and\ \bibinfo
  {author} {\bibfnamefont {A.}~\bibnamefont {Ukawa}},\ }\bibfield  {booktitle}
  {\emph {\bibinfo {booktitle} {{Proceedings, 34th International Symposium on
  Lattice Field Theory (Lattice 2016): Southampton, UK, July 24-30, 2016}}},\
  }\href@noop {} {\bibfield  {journal} {\bibinfo  {journal} {PoS}\ }\textbf
  {\bibinfo {volume} {LATTICE2016}},\ \bibinfo {pages} {384} (\bibinfo {year}
  {2017})},\ \Eprint {http://arxiv.org/abs/1612.05371} {arXiv:1612.05371
  [hep-lat]} \BibitemShut {NoStop}%
\bibitem [{\citenamefont {Bazavov}\ \emph
  {et~al.}(2017{\natexlab{b}})\citenamefont {Bazavov}, \citenamefont {Ding},
  \citenamefont {Hegde}, \citenamefont {Karsch}, \citenamefont {Laermann},
  \citenamefont {Mukherjee}, \citenamefont {Petreczky},\ and\ \citenamefont
  {Schmidt}}]{Bazavov:2017xul}%
  \BibitemOpen
  \bibfield  {author} {\bibinfo {author} {\bibfnamefont {A.}~\bibnamefont
  {Bazavov}}, \bibinfo {author} {\bibfnamefont {H.~T.}\ \bibnamefont {Ding}},
  \bibinfo {author} {\bibfnamefont {P.}~\bibnamefont {Hegde}}, \bibinfo
  {author} {\bibfnamefont {F.}~\bibnamefont {Karsch}}, \bibinfo {author}
  {\bibfnamefont {E.}~\bibnamefont {Laermann}}, \bibinfo {author}
  {\bibfnamefont {S.}~\bibnamefont {Mukherjee}}, \bibinfo {author}
  {\bibfnamefont {P.}~\bibnamefont {Petreczky}}, \ and\ \bibinfo {author}
  {\bibfnamefont {C.}~\bibnamefont {Schmidt}},\ }\href {\doibase
  10.1103/PhysRevD.95.074505} {\bibfield  {journal} {\bibinfo  {journal} {Phys.
  Rev.}\ }\textbf {\bibinfo {volume} {D95}},\ \bibinfo {pages} {074505}
  (\bibinfo {year} {2017}{\natexlab{b}})},\ \Eprint
  {http://arxiv.org/abs/1701.03548} {arXiv:1701.03548 [hep-lat]} \BibitemShut
  {NoStop}%
\bibitem [{\citenamefont {Cuteri}\ \emph
  {et~al.}(2021{\natexlab{b}})\citenamefont {Cuteri}, \citenamefont
  {Philipsen},\ and\ \citenamefont {Sciarra}}]{Cuteri:2021ikv}%
  \BibitemOpen
  \bibfield  {author} {\bibinfo {author} {\bibfnamefont {F.}~\bibnamefont
  {Cuteri}}, \bibinfo {author} {\bibfnamefont {O.}~\bibnamefont {Philipsen}}, \
  and\ \bibinfo {author} {\bibfnamefont {A.}~\bibnamefont {Sciarra}},\ }\href
  {\doibase 10.1007/JHEP11(2021)141} {\bibfield  {journal} {\bibinfo  {journal}
  {JHEP}\ }\textbf {\bibinfo {volume} {11}},\ \bibinfo {pages} {141} (\bibinfo
  {year} {2021}{\natexlab{b}})},\ \Eprint {http://arxiv.org/abs/2107.12739}
  {arXiv:2107.12739 [hep-lat]} \BibitemShut {NoStop}%
\bibitem [{\citenamefont {Dini}\ \emph {et~al.}(2022)\citenamefont {Dini},
  \citenamefont {Hegde}, \citenamefont {Karsch}, \citenamefont {Lahiri},
  \citenamefont {Schmidt},\ and\ \citenamefont {Sharma}}]{Dini:2021hug}%
  \BibitemOpen
  \bibfield  {author} {\bibinfo {author} {\bibfnamefont {L.}~\bibnamefont
  {Dini}}, \bibinfo {author} {\bibfnamefont {P.}~\bibnamefont {Hegde}},
  \bibinfo {author} {\bibfnamefont {F.}~\bibnamefont {Karsch}}, \bibinfo
  {author} {\bibfnamefont {A.}~\bibnamefont {Lahiri}}, \bibinfo {author}
  {\bibfnamefont {C.}~\bibnamefont {Schmidt}}, \ and\ \bibinfo {author}
  {\bibfnamefont {S.}~\bibnamefont {Sharma}},\ }\href {\doibase
  10.1103/PhysRevD.105.034510} {\bibfield  {journal} {\bibinfo  {journal}
  {Phys. Rev. D}\ }\textbf {\bibinfo {volume} {105}},\ \bibinfo {pages}
  {034510} (\bibinfo {year} {2022})},\ \Eprint
  {http://arxiv.org/abs/2111.12599} {arXiv:2111.12599 [hep-lat]} \BibitemShut
  {NoStop}%
\bibitem [{\citenamefont {Fejos}(2022)}]{Fejos:2022mso}%
  \BibitemOpen
  \bibfield  {author} {\bibinfo {author} {\bibfnamefont {G.}~\bibnamefont
  {Fejos}},\ }\href {\doibase 10.1103/PhysRevD.105.L071506} {\bibfield
  {journal} {\bibinfo  {journal} {Phys. Rev. D}\ }\textbf {\bibinfo {volume}
  {105}},\ \bibinfo {pages} {L071506} (\bibinfo {year} {2022})},\ \Eprint
  {http://arxiv.org/abs/2201.07909} {arXiv:2201.07909 [hep-ph]} \BibitemShut
  {NoStop}%
\bibitem [{\citenamefont {Fischer}\ \emph {et~al.}(2007)\citenamefont
  {Fischer}, \citenamefont {Nickel},\ and\ \citenamefont
  {Wambach}}]{Fischer:2007ze}%
  \BibitemOpen
  \bibfield  {author} {\bibinfo {author} {\bibfnamefont {C.~S.}\ \bibnamefont
  {Fischer}}, \bibinfo {author} {\bibfnamefont {D.}~\bibnamefont {Nickel}}, \
  and\ \bibinfo {author} {\bibfnamefont {J.}~\bibnamefont {Wambach}},\ }\href
  {\doibase 10.1103/PhysRevD.76.094009} {\bibfield  {journal} {\bibinfo
  {journal} {Phys. Rev. D}\ }\textbf {\bibinfo {volume} {76}},\ \bibinfo
  {pages} {094009} (\bibinfo {year} {2007})},\ \Eprint
  {http://arxiv.org/abs/0705.4407} {arXiv:0705.4407 [hep-ph]} \BibitemShut
  {NoStop}%
\bibitem [{\citenamefont {Fischer}\ \emph {et~al.}(2009)\citenamefont
  {Fischer}, \citenamefont {Nickel},\ and\ \citenamefont
  {Williams}}]{Fischer:2008sp}%
  \BibitemOpen
  \bibfield  {author} {\bibinfo {author} {\bibfnamefont {C.~S.}\ \bibnamefont
  {Fischer}}, \bibinfo {author} {\bibfnamefont {D.}~\bibnamefont {Nickel}}, \
  and\ \bibinfo {author} {\bibfnamefont {R.}~\bibnamefont {Williams}},\ }\href
  {\doibase 10.1140/epjc/s10052-008-0821-1} {\bibfield  {journal} {\bibinfo
  {journal} {Eur. Phys. J. C}\ }\textbf {\bibinfo {volume} {60}},\ \bibinfo
  {pages} {47} (\bibinfo {year} {2009})},\ \Eprint
  {http://arxiv.org/abs/0807.3486} {arXiv:0807.3486 [hep-ph]} \BibitemShut
  {NoStop}%
\bibitem [{\citenamefont {Gunkel}\ and\ \citenamefont
  {Fischer}(2021{\natexlab{b}})}]{Gunkel:2020wcl}%
  \BibitemOpen
  \bibfield  {author} {\bibinfo {author} {\bibfnamefont {P.~J.}\ \bibnamefont
  {Gunkel}}\ and\ \bibinfo {author} {\bibfnamefont {C.~S.}\ \bibnamefont
  {Fischer}},\ }\href {\doibase 10.1140/epja/s10050-021-00450-7} {\bibfield
  {journal} {\bibinfo  {journal} {Eur. Phys. J. A}\ }\textbf {\bibinfo {volume}
  {57}},\ \bibinfo {pages} {147} (\bibinfo {year} {2021}{\natexlab{b}})},\
  \Eprint {http://arxiv.org/abs/2012.01957} {arXiv:2012.01957 [hep-ph]}
  \BibitemShut {NoStop}%
\bibitem [{\citenamefont {Gunkel}\ \emph {et~al.}(2019)\citenamefont {Gunkel},
  \citenamefont {Fischer},\ and\ \citenamefont {Isserstedt}}]{Gunkel:2019xnh}%
  \BibitemOpen
  \bibfield  {author} {\bibinfo {author} {\bibfnamefont {P.~J.}\ \bibnamefont
  {Gunkel}}, \bibinfo {author} {\bibfnamefont {C.~S.}\ \bibnamefont {Fischer}},
  \ and\ \bibinfo {author} {\bibfnamefont {P.}~\bibnamefont {Isserstedt}},\
  }\href {\doibase 10.1140/epja/i2019-12868-1} {\bibfield  {journal} {\bibinfo
  {journal} {Eur. Phys. J. A}\ }\textbf {\bibinfo {volume} {55}},\ \bibinfo
  {pages} {169} (\bibinfo {year} {2019})},\ \Eprint
  {http://arxiv.org/abs/1907.08110} {arXiv:1907.08110 [hep-ph]} \BibitemShut
  {NoStop}%
\bibitem [{\citenamefont {Fischer}\ and\ \citenamefont
  {Luecker}(2013)}]{Fischer:2012vc}%
  \BibitemOpen
  \bibfield  {author} {\bibinfo {author} {\bibfnamefont {C.~S.}\ \bibnamefont
  {Fischer}}\ and\ \bibinfo {author} {\bibfnamefont {J.}~\bibnamefont
  {Luecker}},\ }\href {\doibase 10.1016/j.physletb.2012.11.054} {\bibfield
  {journal} {\bibinfo  {journal} {Phys. Lett.}\ }\textbf {\bibinfo {volume}
  {B718}},\ \bibinfo {pages} {1036} (\bibinfo {year} {2013})},\ \Eprint
  {http://arxiv.org/abs/1206.5191} {arXiv:1206.5191 [hep-ph]} \BibitemShut
  {NoStop}%
\bibitem [{\citenamefont {Eichmann}\ \emph
  {et~al.}(2016{\natexlab{a}})\citenamefont {Eichmann}, \citenamefont
  {Fischer},\ and\ \citenamefont {Welzbacher}}]{Eichmann:2015kfa}%
  \BibitemOpen
  \bibfield  {author} {\bibinfo {author} {\bibfnamefont {G.}~\bibnamefont
  {Eichmann}}, \bibinfo {author} {\bibfnamefont {C.~S.}\ \bibnamefont
  {Fischer}}, \ and\ \bibinfo {author} {\bibfnamefont {C.~A.}\ \bibnamefont
  {Welzbacher}},\ }\href {\doibase 10.1103/PhysRevD.93.034013} {\bibfield
  {journal} {\bibinfo  {journal} {Phys. Rev. D}\ }\textbf {\bibinfo {volume}
  {93}},\ \bibinfo {pages} {034013} (\bibinfo {year} {2016}{\natexlab{a}})},\
  \Eprint {http://arxiv.org/abs/1509.02082} {arXiv:1509.02082 [hep-ph]}
  \BibitemShut {NoStop}%
\bibitem [{\citenamefont {Isserstedt}\ \emph {et~al.}(2021)\citenamefont
  {Isserstedt}, \citenamefont {Fischer},\ and\ \citenamefont
  {Steinert}}]{Isserstedt:2020qll}%
  \BibitemOpen
  \bibfield  {author} {\bibinfo {author} {\bibfnamefont {P.}~\bibnamefont
  {Isserstedt}}, \bibinfo {author} {\bibfnamefont {C.~S.}\ \bibnamefont
  {Fischer}}, \ and\ \bibinfo {author} {\bibfnamefont {T.}~\bibnamefont
  {Steinert}},\ }\href {\doibase 10.1103/PhysRevD.103.054012} {\bibfield
  {journal} {\bibinfo  {journal} {Phys. Rev. D}\ }\textbf {\bibinfo {volume}
  {103}},\ \bibinfo {pages} {054012} (\bibinfo {year} {2021})},\ \Eprint
  {http://arxiv.org/abs/2012.04991} {arXiv:2012.04991 [hep-ph]} \BibitemShut
  {NoStop}%
\bibitem [{\citenamefont {Fischer}\ and\ \citenamefont
  {Williams}(2008)}]{Fischer:2008wy}%
  \BibitemOpen
  \bibfield  {author} {\bibinfo {author} {\bibfnamefont {C.~S.}\ \bibnamefont
  {Fischer}}\ and\ \bibinfo {author} {\bibfnamefont {R.}~\bibnamefont
  {Williams}},\ }\href {\doibase 10.1103/PhysRevD.78.074006} {\bibfield
  {journal} {\bibinfo  {journal} {Phys. Rev. D}\ }\textbf {\bibinfo {volume}
  {78}},\ \bibinfo {pages} {074006} (\bibinfo {year} {2008})},\ \Eprint
  {http://arxiv.org/abs/0808.3372} {arXiv:0808.3372 [hep-ph]} \BibitemShut
  {NoStop}%
\bibitem [{\citenamefont {Sanchis-Alepuz}\ \emph {et~al.}(2014)\citenamefont
  {Sanchis-Alepuz}, \citenamefont {Fischer},\ and\ \citenamefont
  {Kubrak}}]{Sanchis-Alepuz:2014wea}%
  \BibitemOpen
  \bibfield  {author} {\bibinfo {author} {\bibfnamefont {H.}~\bibnamefont
  {Sanchis-Alepuz}}, \bibinfo {author} {\bibfnamefont {C.~S.}\ \bibnamefont
  {Fischer}}, \ and\ \bibinfo {author} {\bibfnamefont {S.}~\bibnamefont
  {Kubrak}},\ }\href {\doibase 10.1016/j.physletb.2014.04.031} {\bibfield
  {journal} {\bibinfo  {journal} {Phys. Lett.}\ }\textbf {\bibinfo {volume}
  {B733}},\ \bibinfo {pages} {151} (\bibinfo {year} {2014})},\ \Eprint
  {http://arxiv.org/abs/1401.3183} {arXiv:1401.3183 [hep-ph]} \BibitemShut
  {NoStop}%
\bibitem [{\citenamefont {Ball}\ and\ \citenamefont
  {Chiu}(1980)}]{Ball:1980ay}%
  \BibitemOpen
  \bibfield  {author} {\bibinfo {author} {\bibfnamefont {J.~S.}\ \bibnamefont
  {Ball}}\ and\ \bibinfo {author} {\bibfnamefont {T.-W.}\ \bibnamefont
  {Chiu}},\ }\href {\doibase 10.1103/PhysRevD.22.2542} {\bibfield  {journal}
  {\bibinfo  {journal} {Phys. Rev. D}\ }\textbf {\bibinfo {volume} {22}},\
  \bibinfo {pages} {2542} (\bibinfo {year} {1980})}\BibitemShut {NoStop}%
\bibitem [{\citenamefont {Fischer}\ and\ \citenamefont
  {Mueller}(2011)}]{Fischer:2011pk}%
  \BibitemOpen
  \bibfield  {author} {\bibinfo {author} {\bibfnamefont {C.~S.}\ \bibnamefont
  {Fischer}}\ and\ \bibinfo {author} {\bibfnamefont {J.~A.}\ \bibnamefont
  {Mueller}},\ }\href {\doibase 10.1103/PhysRevD.84.054013} {\bibfield
  {journal} {\bibinfo  {journal} {Phys. Rev. D}\ }\textbf {\bibinfo {volume}
  {84}},\ \bibinfo {pages} {054013} (\bibinfo {year} {2011})},\ \Eprint
  {http://arxiv.org/abs/1106.2700} {arXiv:1106.2700 [hep-ph]} \BibitemShut
  {NoStop}%
\bibitem [{\citenamefont {Fischer}\ \emph {et~al.}(2010)\citenamefont
  {Fischer}, \citenamefont {Maas},\ and\ \citenamefont
  {Mueller}}]{Fischer:2010fx}%
  \BibitemOpen
  \bibfield  {author} {\bibinfo {author} {\bibfnamefont {C.~S.}\ \bibnamefont
  {Fischer}}, \bibinfo {author} {\bibfnamefont {A.}~\bibnamefont {Maas}}, \
  and\ \bibinfo {author} {\bibfnamefont {J.~A.}\ \bibnamefont {Mueller}},\
  }\href {\doibase 10.1140/epjc/s10052-010-1343-1} {\bibfield  {journal}
  {\bibinfo  {journal} {Eur. Phys. J. C}\ }\textbf {\bibinfo {volume} {68}},\
  \bibinfo {pages} {165} (\bibinfo {year} {2010})},\ \Eprint
  {http://arxiv.org/abs/1003.1960} {arXiv:1003.1960 [hep-ph]} \BibitemShut
  {NoStop}%
\bibitem [{\citenamefont {Maas}\ \emph {et~al.}(2012)\citenamefont {Maas},
  \citenamefont {Pawlowski}, \citenamefont {{von Smekal}},\ and\ \citenamefont
  {Spielmann}}]{Maas:2011ez}%
  \BibitemOpen
  \bibfield  {author} {\bibinfo {author} {\bibfnamefont {A.}~\bibnamefont
  {Maas}}, \bibinfo {author} {\bibfnamefont {J.~M.}\ \bibnamefont {Pawlowski}},
  \bibinfo {author} {\bibfnamefont {L.}~\bibnamefont {{von Smekal}}}, \ and\
  \bibinfo {author} {\bibfnamefont {D.}~\bibnamefont {Spielmann}},\ }\href
  {\doibase 10.1103/PhysRevD.85.034037} {\bibfield  {journal} {\bibinfo
  {journal} {Phys. Rev. D}\ }\textbf {\bibinfo {volume} {85}},\ \bibinfo
  {pages} {034037} (\bibinfo {year} {2012})},\ \Eprint
  {http://arxiv.org/abs/1110.6340} {arXiv:1110.6340 [hep-lat]} \BibitemShut
  {NoStop}%
\bibitem [{\citenamefont {Heupel}\ \emph {et~al.}(2014)\citenamefont {Heupel},
  \citenamefont {Goecke},\ and\ \citenamefont {Fischer}}]{Heupel:2014ina}%
  \BibitemOpen
  \bibfield  {author} {\bibinfo {author} {\bibfnamefont {W.}~\bibnamefont
  {Heupel}}, \bibinfo {author} {\bibfnamefont {T.}~\bibnamefont {Goecke}}, \
  and\ \bibinfo {author} {\bibfnamefont {C.~S.}\ \bibnamefont {Fischer}},\
  }\href {\doibase 10.1140/epja/i2014-14085-x} {\bibfield  {journal} {\bibinfo
  {journal} {Eur. Phys. J. A}\ }\textbf {\bibinfo {volume} {50}},\ \bibinfo
  {pages} {85} (\bibinfo {year} {2014})},\ \Eprint
  {http://arxiv.org/abs/1402.5042} {arXiv:1402.5042 [hep-ph]} \BibitemShut
  {NoStop}%
\bibitem [{\citenamefont {Son}\ and\ \citenamefont
  {Stephanov}(2002)}]{Son:2001ff}%
  \BibitemOpen
  \bibfield  {author} {\bibinfo {author} {\bibfnamefont {D.~T.}\ \bibnamefont
  {Son}}\ and\ \bibinfo {author} {\bibfnamefont {M.~A.}\ \bibnamefont
  {Stephanov}},\ }\href {\doibase 10.1103/PhysRevLett.88.202302} {\bibfield
  {journal} {\bibinfo  {journal} {Phys. Rev. Lett.}\ }\textbf {\bibinfo
  {volume} {88}},\ \bibinfo {pages} {202302} (\bibinfo {year} {2002})},\
  \Eprint {http://arxiv.org/abs/hep-ph/0111100} {arXiv:hep-ph/0111100 [hep-ph]}
  \BibitemShut {NoStop}%
\bibitem [{\citenamefont {Maris}\ \emph {et~al.}(2001)\citenamefont {Maris},
  \citenamefont {Roberts}, \citenamefont {Schmidt},\ and\ \citenamefont
  {Tandy}}]{Maris:2000ig}%
  \BibitemOpen
  \bibfield  {author} {\bibinfo {author} {\bibfnamefont {P.}~\bibnamefont
  {Maris}}, \bibinfo {author} {\bibfnamefont {C.~D.}\ \bibnamefont {Roberts}},
  \bibinfo {author} {\bibfnamefont {S.~M.}\ \bibnamefont {Schmidt}}, \ and\
  \bibinfo {author} {\bibfnamefont {P.~C.}\ \bibnamefont {Tandy}},\ }\href
  {\doibase 10.1103/PhysRevC.63.025202} {\bibfield  {journal} {\bibinfo
  {journal} {Phys. Rev.}\ }\textbf {\bibinfo {volume} {C63}},\ \bibinfo {pages}
  {025202} (\bibinfo {year} {2001})},\ \Eprint
  {http://arxiv.org/abs/nucl-th/0001064} {arXiv:nucl-th/0001064 [nucl-th]}
  \BibitemShut {NoStop}%
\bibitem [{\citenamefont {Maris}\ and\ \citenamefont
  {Roberts}(1997)}]{Maris:1997tm}%
  \BibitemOpen
  \bibfield  {author} {\bibinfo {author} {\bibfnamefont {P.}~\bibnamefont
  {Maris}}\ and\ \bibinfo {author} {\bibfnamefont {C.~D.}\ \bibnamefont
  {Roberts}},\ }\href {\doibase 10.1103/PhysRevC.56.3369} {\bibfield  {journal}
  {\bibinfo  {journal} {Phys. Rev.}\ }\textbf {\bibinfo {volume} {C56}},\
  \bibinfo {pages} {3369} (\bibinfo {year} {1997})},\ \Eprint
  {http://arxiv.org/abs/nucl-th/9708029} {arXiv:nucl-th/9708029 [nucl-th]}
  \BibitemShut {NoStop}%
\bibitem [{\citenamefont {Maris}\ \emph {et~al.}(1998)\citenamefont {Maris},
  \citenamefont {Roberts},\ and\ \citenamefont {Tandy}}]{Maris:1997hd}%
  \BibitemOpen
  \bibfield  {author} {\bibinfo {author} {\bibfnamefont {P.}~\bibnamefont
  {Maris}}, \bibinfo {author} {\bibfnamefont {C.~D.}\ \bibnamefont {Roberts}},
  \ and\ \bibinfo {author} {\bibfnamefont {P.~C.}\ \bibnamefont {Tandy}},\
  }\href {\doibase 10.1016/S0370-2693(97)01535-9} {\bibfield  {journal}
  {\bibinfo  {journal} {Phys. Lett.}\ }\textbf {\bibinfo {volume} {B420}},\
  \bibinfo {pages} {267} (\bibinfo {year} {1998})},\ \Eprint
  {http://arxiv.org/abs/nucl-th/9707003} {arXiv:nucl-th/9707003 [nucl-th]}
  \BibitemShut {NoStop}%
\bibitem [{\citenamefont {Eichmann}\ \emph
  {et~al.}(2016{\natexlab{b}})\citenamefont {Eichmann}, \citenamefont
  {Sanchis-Alepuz}, \citenamefont {Williams}, \citenamefont {Alkofer},\ and\
  \citenamefont {Fischer}}]{Eichmann:2016yit}%
  \BibitemOpen
  \bibfield  {author} {\bibinfo {author} {\bibfnamefont {G.}~\bibnamefont
  {Eichmann}}, \bibinfo {author} {\bibfnamefont {H.}~\bibnamefont
  {Sanchis-Alepuz}}, \bibinfo {author} {\bibfnamefont {R.}~\bibnamefont
  {Williams}}, \bibinfo {author} {\bibfnamefont {R.}~\bibnamefont {Alkofer}}, \
  and\ \bibinfo {author} {\bibfnamefont {C.~S.}\ \bibnamefont {Fischer}},\
  }\href {\doibase 10.1016/j.ppnp.2016.07.001} {\bibfield  {journal} {\bibinfo
  {journal} {Prog. Part. Nucl. Phys.}\ }\textbf {\bibinfo {volume} {91}},\
  \bibinfo {pages} {1} (\bibinfo {year} {2016}{\natexlab{b}})},\ \Eprint
  {http://arxiv.org/abs/1606.09602} {arXiv:1606.09602 [hep-ph]} \BibitemShut
  {NoStop}%
\bibitem [{\citenamefont {L\"ucker}(2013)}]{Lucker:2013uya}%
  \BibitemOpen
  \bibfield  {author} {\bibinfo {author} {\bibfnamefont {J.}~\bibnamefont
  {L\"ucker}},\ }\emph {\bibinfo {title} {{Chiral and deconfinement phase
  transitions in $N{_f}=2$ and $N{_f}=2+1$ quantum chromodynamics}}},\
  \href@noop {} {Ph.D. thesis},\ \bibinfo  {school} {Giessen U.} (\bibinfo
  {year} {2013})\BibitemShut {NoStop}%
\bibitem [{\citenamefont {Pagels}\ and\ \citenamefont
  {Stokar}(1979)}]{Pagels:1979hd}%
  \BibitemOpen
  \bibfield  {author} {\bibinfo {author} {\bibfnamefont {H.}~\bibnamefont
  {Pagels}}\ and\ \bibinfo {author} {\bibfnamefont {S.}~\bibnamefont
  {Stokar}},\ }\href {\doibase 10.1103/PhysRevD.20.2947} {\bibfield  {journal}
  {\bibinfo  {journal} {Phys. Rev. D}\ }\textbf {\bibinfo {volume} {20}},\
  \bibinfo {pages} {2947} (\bibinfo {year} {1979})}\BibitemShut {NoStop}%
\bibitem [{\citenamefont {Ding}\ \emph {et~al.}(2019)\citenamefont {Ding} \emph
  {et~al.}}]{HotQCD:2019xnw}%
  \BibitemOpen
  \bibfield  {author} {\bibinfo {author} {\bibfnamefont {H.~T.}\ \bibnamefont
  {Ding}} \emph {et~al.} (\bibinfo {collaboration} {HotQCD}),\ }\href {\doibase
  10.1103/PhysRevLett.123.062002} {\bibfield  {journal} {\bibinfo  {journal}
  {Phys. Rev. Lett.}\ }\textbf {\bibinfo {volume} {123}},\ \bibinfo {pages}
  {062002} (\bibinfo {year} {2019})},\ \Eprint
  {http://arxiv.org/abs/1903.04801} {arXiv:1903.04801 [hep-lat]} \BibitemShut
  {NoStop}%
\bibitem [{\citenamefont {Gao}\ and\ \citenamefont
  {Pawlowski}(2021{\natexlab{b}})}]{Gao:2021vsf}%
  \BibitemOpen
  \bibfield  {author} {\bibinfo {author} {\bibfnamefont {F.}~\bibnamefont
  {Gao}}\ and\ \bibinfo {author} {\bibfnamefont {J.~M.}\ \bibnamefont
  {Pawlowski}},\ }\href {\doibase 10.1103/PhysRevD.105.094020} {\bibfield
  {journal} {\bibinfo  {journal} {Phys. Rev. D}\ }\textbf {\bibinfo {volume}
  {105}},\ \bibinfo {pages} {094020} (\bibinfo {year} {2021}{\natexlab{b}})},\
  \Eprint {http://arxiv.org/abs/2112.01395} {arXiv:2112.01395 [hep-ph]}
  \BibitemShut {NoStop}%
\bibitem [{\citenamefont {Bornyakov}\ \emph {et~al.}(2010)\citenamefont
  {Bornyakov}, \citenamefont {Horsley}, \citenamefont {Morozov}, \citenamefont
  {Nakamura}, \citenamefont {Polikarpov}, \citenamefont {Rakow}, \citenamefont
  {Schierholz},\ and\ \citenamefont {Suzuki}}]{Bornyakov:2009qh}%
  \BibitemOpen
  \bibfield  {author} {\bibinfo {author} {\bibfnamefont {V.~G.}\ \bibnamefont
  {Bornyakov}}, \bibinfo {author} {\bibfnamefont {R.}~\bibnamefont {Horsley}},
  \bibinfo {author} {\bibfnamefont {S.~M.}\ \bibnamefont {Morozov}}, \bibinfo
  {author} {\bibfnamefont {Y.}~\bibnamefont {Nakamura}}, \bibinfo {author}
  {\bibfnamefont {M.~I.}\ \bibnamefont {Polikarpov}}, \bibinfo {author}
  {\bibfnamefont {P.~E.~L.}\ \bibnamefont {Rakow}}, \bibinfo {author}
  {\bibfnamefont {G.}~\bibnamefont {Schierholz}}, \ and\ \bibinfo {author}
  {\bibfnamefont {T.}~\bibnamefont {Suzuki}},\ }\href {\doibase
  10.1103/PhysRevD.82.014504} {\bibfield  {journal} {\bibinfo  {journal} {Phys.
  Rev. D}\ }\textbf {\bibinfo {volume} {82}},\ \bibinfo {pages} {014504}
  (\bibinfo {year} {2010})},\ \Eprint {http://arxiv.org/abs/0910.2392}
  {arXiv:0910.2392 [hep-lat]} \BibitemShut {NoStop}%
\bibitem [{\citenamefont {Roberts}\ and\ \citenamefont
  {Schmidt}(2000)}]{Roberts:2000aa}%
  \BibitemOpen
  \bibfield  {author} {\bibinfo {author} {\bibfnamefont {C.~D.}\ \bibnamefont
  {Roberts}}\ and\ \bibinfo {author} {\bibfnamefont {S.~M.}\ \bibnamefont
  {Schmidt}},\ }\href {\doibase 10.1016/S0146-6410(00)90011-5} {\bibfield
  {journal} {\bibinfo  {journal} {Prog. Part. Nucl. Phys.}\ }\textbf {\bibinfo
  {volume} {45}},\ \bibinfo {pages} {S1} (\bibinfo {year} {2000})},\ \Eprint
  {http://arxiv.org/abs/nucl-th/0005064} {arXiv:nucl-th/0005064} \BibitemShut
  {NoStop}%
\bibitem [{\citenamefont {Berges}\ \emph {et~al.}(2002)\citenamefont {Berges},
  \citenamefont {Tetradis},\ and\ \citenamefont {Wetterich}}]{Berges:2000ew}%
  \BibitemOpen
  \bibfield  {author} {\bibinfo {author} {\bibfnamefont {J.}~\bibnamefont
  {Berges}}, \bibinfo {author} {\bibfnamefont {N.}~\bibnamefont {Tetradis}}, \
  and\ \bibinfo {author} {\bibfnamefont {C.}~\bibnamefont {Wetterich}},\ }\href
  {\doibase 10.1016/S0370-1573(01)00098-9} {\bibfield  {journal} {\bibinfo
  {journal} {Phys. Rept.}\ }\textbf {\bibinfo {volume} {363}},\ \bibinfo
  {pages} {223} (\bibinfo {year} {2002})},\ \Eprint
  {http://arxiv.org/abs/hep-ph/0005122} {arXiv:hep-ph/0005122 [hep-ph]}
  \BibitemShut {NoStop}%
\bibitem [{\citenamefont {Schaefer}\ and\ \citenamefont
  {Wambach}(2005)}]{Schaefer:2004en}%
  \BibitemOpen
  \bibfield  {author} {\bibinfo {author} {\bibfnamefont {B.-J.}\ \bibnamefont
  {Schaefer}}\ and\ \bibinfo {author} {\bibfnamefont {J.}~\bibnamefont
  {Wambach}},\ }\href {\doibase 10.1016/j.nuclphysa.2005.04.012} {\bibfield
  {journal} {\bibinfo  {journal} {Nucl. Phys. A}\ }\textbf {\bibinfo {volume}
  {757}},\ \bibinfo {pages} {479} (\bibinfo {year} {2005})},\ \Eprint
  {http://arxiv.org/abs/nucl-th/0403039} {arXiv:nucl-th/0403039} \BibitemShut
  {NoStop}%
\bibitem [{\citenamefont {Schaefer}\ and\ \citenamefont
  {Wambach}(2007)}]{Schaefer:2006ds}%
  \BibitemOpen
  \bibfield  {author} {\bibinfo {author} {\bibfnamefont {B.-J.}\ \bibnamefont
  {Schaefer}}\ and\ \bibinfo {author} {\bibfnamefont {J.}~\bibnamefont
  {Wambach}},\ }\href {\doibase 10.1103/PhysRevD.75.085015} {\bibfield
  {journal} {\bibinfo  {journal} {Phys. Rev.}\ }\textbf {\bibinfo {volume}
  {D75}},\ \bibinfo {pages} {085015} (\bibinfo {year} {2007})},\ \Eprint
  {http://arxiv.org/abs/hep-ph/0603256} {arXiv:hep-ph/0603256 [hep-ph]}
  \BibitemShut {NoStop}%
\bibitem [{\citenamefont {Rennecke}\ and\ \citenamefont
  {Schaefer}(2017)}]{Rennecke:2016tkm}%
  \BibitemOpen
  \bibfield  {author} {\bibinfo {author} {\bibfnamefont {F.}~\bibnamefont
  {Rennecke}}\ and\ \bibinfo {author} {\bibfnamefont {B.-J.}\ \bibnamefont
  {Schaefer}},\ }\href {\doibase 10.1103/PhysRevD.96.016009} {\bibfield
  {journal} {\bibinfo  {journal} {Phys. Rev.}\ }\textbf {\bibinfo {volume}
  {D96}},\ \bibinfo {pages} {016009} (\bibinfo {year} {2017})},\ \Eprint
  {http://arxiv.org/abs/1610.08748} {arXiv:1610.08748 [hep-ph]} \BibitemShut
  {NoStop}%
\bibitem [{\citenamefont {D'Ambrosio}\ \emph {et~al.}(2023)\citenamefont
  {D'Ambrosio}, \citenamefont {Philipsen},\ and\ \citenamefont
  {Kaiser}}]{DAmbrosio:2022kig}%
  \BibitemOpen
  \bibfield  {author} {\bibinfo {author} {\bibfnamefont {A.}~\bibnamefont
  {D'Ambrosio}}, \bibinfo {author} {\bibfnamefont {O.}~\bibnamefont
  {Philipsen}}, \ and\ \bibinfo {author} {\bibfnamefont {R.}~\bibnamefont
  {Kaiser}},\ }\href {\doibase 10.22323/1.430.0172} {\bibfield  {journal}
  {\bibinfo  {journal} {PoS}\ }\textbf {\bibinfo {volume} {LATTICE2022}},\
  \bibinfo {pages} {172} (\bibinfo {year} {2023})},\ \Eprint
  {http://arxiv.org/abs/2212.03655} {arXiv:2212.03655 [hep-lat]} \BibitemShut
  {NoStop}%
\bibitem [{\citenamefont {Ellis}(2017)}]{Ellis:2016jkw}%
  \BibitemOpen
  \bibfield  {author} {\bibinfo {author} {\bibfnamefont {J.}~\bibnamefont
  {Ellis}},\ }\href {\doibase 10.1016/j.cpc.2016.08.019} {\bibfield  {journal}
  {\bibinfo  {journal} {Comput. Phys. Commun.}\ }\textbf {\bibinfo {volume}
  {210}},\ \bibinfo {pages} {103} (\bibinfo {year} {2017})},\ \Eprint
  {http://arxiv.org/abs/1601.05437} {arXiv:1601.05437 [hep-ph]} \BibitemShut
  {NoStop}%
\end{thebibliography}%

\end{document}